\documentclass{article}

\usepackage{amsfonts}
\usepackage{epsfig}

\begin{document}

\title{Random Walks Along the Streets and Canals in Compact Cities: Spectral analysis, Dynamical
Modularity, Information, and Statistical
Mechanics}

\vspace{1cm}

\author{ {D. Volchenkov} \footnote{The Alexander von Humboldt Research Fellow at the
BiBoS Research Center}  and {Ph. Blanchard}
\vspace{0.5cm}\\
{\it  BiBoS, University Bielefeld, Postfach 100131,}\\
{\it D-33501, Bielefeld, Germany} \\
{\it Phone: +49 (0)521 / 106-2972 } \\
{\it Fax: +49 (0)521 / 106-6455 } \\
{\it E-Mail: VOLCHENK@Physik.Uni-Bielefeld.DE}}

\date{\today}
\maketitle

\begin{abstract}
Different models of random walks on the dual graphs of compact
urban structures are considered. Analysis of access times between
streets helps to detect the city modularity. The statistical
mechanics approach to the ensembles of lazy random walkers is
developed. The complexity of city modularity can be measured by an
information-like parameter which plays the role of an individual
fingerprint of {\it Genius loci}.
 Global structural properties of a city
can be characterized by the thermodynamical parameters calculated
in the random walks problem.
\end{abstract}

\vspace{0.5cm}

\leftline{\textbf{ PACS codes: 89.75.Fb, 89.75.-k, 89.90.+n} }
 \vspace{0.5cm}

\leftline{\textbf{ Keywords: Random walks, complex networks, city streets and canals} }

\section{Introduction to city networks studies}
 \noindent

Studies of urban networks have a long history. Many researches
have been devoted to the optimizations of transportation routes
and power grids, to the predictions of traffic flows between the
highly populated city districts, to the investigations of habits
and artefact exchanges between distanced settlements in historical
eras. In the most of them, relations between different components
of urban structure are often measured along streets, the routes
between junctions which form the nodes of an equivalent planar
graph. The graph-theoretic principles had been applied in
\cite{Nystuen1961} in order to measure the hierarchy in regional
central place systems, in \cite{Kansky63} to the measurement of
transportation networks. The use of graph-theoretic view and
network analysis of spatial systems in geographic science had been
reviewed in \cite{Hagget69}. It is interesting to mention that
graphs have been widely used to represent the connectivity between
offices in buildings \cite{March71} and to classify various
building types in \cite{Steadman83}.

In all these studies, the traffic end points and junctions had been
treated as nodes, and the routes had been considered as
edges of some planar graphs. Being embedded onto the
geographical and economical landscapes, these planar graphs bare
their multiple fingerprints. Among the main factors featuring
them are the high costs  for the maintenance of long-range connections
and the scarce availability of physical space.
 Spatial networks differ from other complex networks and
call for the alternative approaches to investigate them \cite{Cardillo}.

While studying the motifs and cycles in the complex networks, a
comparative analysis of the original graph and of its randomized
version is used. If a motif is statistically significant, it
appears in the real network much frequently than in the randomized
versions of the graph \cite{NewmanSIAM}. However, in the planar
city street patterns, its randomized version is not of
significance since, first, it is surely a non-planar graph due to
the randomness of edge crossings and, second, the long-range
connections which inevitably present in the random graphs in
abundance are extremely costly in the real cities \cite{Cardillo}.
In \cite{BGSKVDT}, it has been proposed to compare the city street
patterns with the grid-like structures that is indeed useful
essentially for the regular urban development.

It had been formulated in the classical essay of T. Harold Hughes
\cite{Hughes1912} that the accessibility of important city objects
for the vehicle traffic and pedestrians is always the chief factor
in regulating the growth and expansion of the city. A broad,
simple  scheme of main traffic lines gives a sense of
connectedness and unity to the various parts of the city and links
up country and town. In \cite{Hillier, Hillier96}, a significant
correlation between the topological accessibility of streets and
their popularity, micro-criminality, micro-economic vitality and
social liveability had been established. In the traditional
representation of space syntax based on relations between streets
through their junctions, the accessibility or distance is
associated with points or junctions.

We would like to mention that the issues of global connectivity of
finite graphs and accessibility of their nodes are the classical
fields of researches in graph theory. They are studied by means of
certain dynamical processes defined on the graphs. In particular,
in order to reach "obscure" parts of large sets and estimate the
probable access times to them,
 {\it random walks} are often used \cite{Lovasz}. There are a number of
other processes that can be defined on a graph describing various
types of diffusion of a large number of random walkers moving on
the network at discrete time steps \cite{BP}. In all such
processes we have deal with discrete time Markov chains studied in
probability theory. Markov chains have the property that their
time evolution
 behavior depends only upon their current state and the state transition
  properties of the model.

At each time step every walker moves from its current node to one
of the neighboring nodes along a randomly selected link. The
metric distance between the nodes are of no matter for such a
discrete time diffusion process and then the focus of study is
naturally shifted from the original problem of traditional space
syntax to the {\it dual} one based on relations between the
streets which themselves are treated as nodes. The distance
between two streets, in such a representation, is a distance in
the graph-theoretic sense. The dual graphs of a geographic network
comparable in its structure with other complex networks are
irrelevant to neither distance, nor the physical space
constraints. The relations between the traditional
 space syntax representation based on the relations between streets through their junctions
and the dual representation that is a morphological representation of relations
between junctions through their streets was studied in details in \cite{Batty}.
The dual city graphs had been developed and studied within the concept of space
 syntax \cite{Hillier}. The key characteristics in space syntax is that precedence
  is given to linear
features such as streets in contrast to fixed points which approximate locations
 \cite{Hillier96}.
The dual city graphs have been discussed recently in \cite{2}.
 In \cite{RTMS}, they have been called the information city network.

In \cite{JC2004},
while identifying a street over a plurality of routes on the city map,
the "named-street" approach had been used, in
which two different arcs of the original street network are
assigned to the same street ID provided they have the same street name.
The main problem of the approach is
that  the meaning of a street name could vary
from one district or quarter  to another even within the same city.
For instance, the streets in Manhattan
do not meet,  in general, the continuity  principle rather playing the
role of local geographical coordinates.

 Being interested in the statistics of random walks, we generalized
the approach used in \cite{JC2004} in a way to account the
possible discontinuities of streets. Namely, we assign an
individual street ID code to each continuous part of it even if
all of them share the same street name. The dual graph is
constructed by mapping edges coded with the same street ID into
nodes of the dual graph, and intersections among each pair of
edges in the original graph - into edges connecting the
corresponding nodes of
 the dual graph like it has been done in \cite{JC2004}.

In \cite{2}, the Intersection Continuity principle (ICN) different
from our identification approach has been used: two edges forming
the largest convex angle in a crossroad on the city map are
 assigned the highest continuity and therefore are coupled together
acquiring the same street ID. The main
 problem with the ICN principle is that the streets crossing
under the convex angles would artificially exchange their
identifiers that is not crucial for the study of degree
statistics, but makes it difficult to interpret the results on
random walks and detect the dynamical modularity of the city.

 It is also important to mention that the
number of street IDs identified within the ICN principle usually
 exceeds substantially the actual number of street names in a city.
  In \cite{Cardillo,2,EPJB,PRE, chaos}, the degree statistics
and various centrality measures for the data
set of the square mile samples of different world cities had been
investigated. However, the decision on which a square mile would
provide an adequate representation of a city is always
questionable.

In this paper, we use an alternative strategy investigating the
spectral properties of random walks defined on the {\it dual}
graphs of {\it compact} city patterns bounded by the natural
geographical limitations. The reason we consider the compact urban
domains is twofold. First, it allows us to avoid the problem of a
"square mile" and, second, the compact urban domains had been
usually developed "at once", in accordance to certain
architectural principles, their partial redevelopment had been
occasional and rear, so that they can be considered typical.

We  have studied the Markov chains defined on the six {\it
undirected}  dual graphs corresponding to the different street and
canal urban structures. Two of them are situated on islands:
Manhattan (with an almost regular greed-like city plan) and the
network of Venice canals (imprinting the joined effect of natural,
political, and economical factors acted on the network during many
centuries). We have also considered two cities founded shortly
after Crusades and developed within the medieval fortresses:
Rothenburg ob der Tauber (the medieval Bavarian city preserving
its original structure from the XIII-th century) and the Bielefeld
downtown (Altstadt Bielefeld) composed of two different parts: the
old one founded in the XIII-th century and the modern part
subjected to the partial urban redevelopment at the end of XIX-th
century. To supplement the study, we have investigated the canal
network of the city of Amsterdam. Although it is not actually
isolated from the national canal network, it is binding to the
delta of Amstel river forming a dense canal web showing a high
degree of radial symmetry. The scarce of physical space is among
the most important factors determining the structure of compact
urban patterns. The general information on the dual graphs of
compact urban street and canal patterns that we have studied is
given in  Tab.~1. The spectral properties of finite Markov chains
defined on these dual graphs are compared  with those of a model
example: a hypothetical village extended along one principal
street and composed of $N$ blind passes branching off it.

A square mile of the New York city grid and a square mile
pattern of Venice streets array have been discussed recently in
\cite{2,EPJB,PRE,chaos}, however, up to our knowledge, the canal
 patterns have never been subjected to the network
analysis. The navigation efficiency in the Manhattan streets has
been studied in \cite{RTMS}.

It is worth to mention the importance of implemented street
identification principle for the conclusion on the degree
statistics of  dual city graphs. The comparative investigations of
different street patterns performed in \cite{Cardillo,2}
implementing the ICN principle reveal  scale-free degree
distributions for the vertices of dual graphs. However, in
\cite{JC2004} it had been reported that under the street-name
approach the dual graphs exhibit the small-world character, but
 scale-free degree statistics can hardly be recognized. The
results on the probability degree statistics for the dual graphs
of compact urban patterns analyzed in accordance to the street
identification principle that we have described above are
compatible with that of \cite{JC2004}. Compact city patterns do
not provide us with  sufficient data to conclude on the
universality of degree statistics. It is remarkable that the
probability degree distributions for the dual graphs correspondent
to the compact city patterns are broad and have a clearly
expressed maximum and a long right tail. The presence of a
noticeable maximum in the probability degree distributions
indicates that the structures of compact urban patterns are
usually close to a regular one and that there is the most probable
number of junctions an average street has in a given city. The
long right tails of distributions correspond to the highly
connected nodes of dual graphs, just a few "broadways",
embankments, and belt roads crossing many more streets than an
average street in the city.
 To give an example, in Fig.~1, we have displayed the log-log plot of fractions
of streets $n(s)$ via the number of their junctions $s$ in
Manhattan. These number are shown by points and the solid line is
for the relevant cumulative distribution
$P_c(s)=\sum_{s'=s}^{\infty}n(s')$ \cite{NewmanSIAM}.

\section{Random walks on the dual city graphs: the description of models
and the sketch of results}
 \noindent

We consider a connected graph $G=(V,E)$ with $|V|=N$
nodes and $|E|=m$ undirected edges specified by its adjacency matrix $A$
such that $A_{ij}=1$ if the node
$i$ is connected to $j$ and $A_{ij}=0$ otherwise.
A random walk starting at a node $v_0\in V$ and traversing a
sequence of random nodes $\{v_t\}$ as $t=0,1,\ldots$ is a Markov chain
characterized by the matrix of
transition probabilities $T=DA$, in which $D$ is the diagonal matrix of
inverse vertex degrees,
$D_{ij}=k^{-1}_{i}\delta_{ij}$, where $k_i=\mathrm{deg}(i)$, the degree
of vertex $i$ in graph $G$
(the number of junctions a street shares on the city plan). The transition
matrix $T$ meets the
normalization condition,
$\sum_jT_{ij}=1.$ We denote by $\pi^t\in \mathbb{R}^N$ the distribution
of $v_t$ in the Markov chain,
$\pi^t_i=\mathrm{Pr}(v_t=i).$
Then the rule of the walk can be expressed by the simple equation
\begin{equation}
\label{distr01}
\pi^{t+1}=T^\dag\pi^t
\end{equation}
where $T^\dag$ is the transposed transition matrix. In the present
paper, we discuss only the undirected dual city graphs with
symmetric  adjacency matrices, $A_{ij}=A_{ji}$. Nowadays, certain
driving directions are specified for the streets by the traffic
regulation polices, so that the relevant transportation lines
appears to be directed. We do not consider them, limiting our
present study only to the network of {\it pedestrian access}. The
distribution of a current node in the random walk defined by
(\ref{distr01}) on the undirected graphs after $t$ steps tends to
a well defined {\it stationary} distribution $\pi_i=k_i/2m$ (which
is uniform if if the graph is regular) that is a left eigenvector
of the transition matrix $T$, belonging to the largest eigenvalue
1 \cite{Lovasz},\cite{LovWinkl}. The spectrum of problem
 (\ref{distr01}) is contained in the interval  $[-1,1]$.

In Sec.~3, we calculate and analyze the expected number of steps
a random walker starting from
node $i$ makes before node $j$ is visited (the access time) for all pairs
of streets (canals) in the compact urban patterns.  The properties of access
times can be used in order to estimate the accessibility
of certain streets and districts by random walkers starting from the rest of city.
In particular, the access times allows one to introduce the equivalence classes of
nodes and
to obtain  a well-defined ordering of these equivalence classes, independent of any
reference node.
The nodes in the lowest class are difficult to reach but easy to get out of, while
 the nodes in the
highest class are easy to reach but difficult to leave. We also discuss the random
 target
access times and the distributions of mean access times in the compact cities.
The latter characteristics can be used in order to detect ghettos (the groups
 of dynamically isolated
nodes) and estimate the accessibility of certain districts from the streets
located in other parts
of the city.

The problem of random walks (\ref{distr01}) defined on finite
graphs can be related to a diffusion process which describes the
dynamics of a large number of random walkers. The associated
Laplace operator  is more convenient, since its spectrum is
positive. Indeed, the eigenvalues in (\ref{distr01}) can be
negative, so that the spectral moments $\sum_i\lambda_i^{-n}$
could oscillate strongly for large networks. In contrast to it,
the eigenvalues of Laplacian are positive that allows one to study
its spectral properties by the powerful methods of statistical
mechanics, in which various functions $f(\lambda_i)$ defined on
the spectrum $\{\lambda_i\}$ are considered.

The diffusion process is defined by the expectation number of
random walkers $\mathbf{n}\in \mathbb{R}^N$ and described by the
equation
\begin{equation}
\label{02}
\dot{\bf n} = {L}{\bf n},
\end{equation}
in which the scaled Laplacian $L$ (symmetric) is defined by
\begin{equation}
\label{01}
L=\mathbf{1}-D^{-1/2}TD^{1/2}
\end{equation}
where $\mathbf{1}$ is the unit matrix. Let us note that $L$ is
related to the transition matrix $T$ in (\ref{distr01}) by
$T=D^{1/2}(\mathbf{1}-L)D^{-1/2}.$
 All
eigenvalues of $L$ belong to the interval $\lambda_\alpha\in[0,2]$, the
eigenvectors are normalized,
 $|\mathbf{n}^{(\alpha)}|=1,$ and orthogonal, $\left\langle\mathbf{n}^{(\alpha)}
\mathbf{n}^{(\beta)} \right\rangle =\delta_{\alpha\beta}$.

The linear equation (\ref{02}) supplied with an initial condition ${\mathbf n}_0$
is solved by the function ${\mathbf n}^t=Q\exp(-tJ)(Q^{-1}{\mathbf n}_0)$, in which
$J$ is a block diagonal matrix (the Jordan canonical form of $L$), $Q$ is the the
transformation matrix corresponding to the Jordan form. Given the eigenvalue
$\lambda_{\alpha}$  with the multiplicity $m_\alpha>1$, the relevant contribution to
${\mathbf n}^t$ is given by
\begin{equation}
\label{03}
\exp(-t\lambda_\alpha)
\sum_{k=0}^{m_\alpha}\left(\sum_{l=0}^{m_\alpha-1}c_{l+1}\frac{t^l}{l!}\right){\mathbf u}_k
\end{equation}
where ${\mathbf u}_k$ is the $k$-th vector of the
$m_\alpha$-dimensional eigenspace of degenerate eigenvalue
$\lambda_\alpha$ and $c_l$ is the $l-$th component of the
transformed vector of initial conditions, ${\mathbf
c}=Q^{-1}{\mathbf n}_0$. It follows from (\ref{03}) that in the
presence of degenerate modes with $m_\alpha\gg 1$ the relaxation
process can be delayed significantly. Moreover, if
$\lambda_\alpha\leq 1$ the expectation  numbers of walkers in a
certain city modules would substantially increase in a lapse of
time.

The analysis of spectral properties of Laplacian operator allows
for detection of fine-scale dynamical modularity of city which
cannot be seen from the transition matrix $T$ and provide us a
tool for the estimation of entire city stability with respect to
occasional cuts of certain transportation lines breaking it up
into dynamically isolated components. In Sec.~4, we consider the
modes of diffusion process defined on the dual city graphs and
compute the coefficients of linear correlations between densities
of random walkers that flow along the edges of dual city graphs in
the discrete steps. Specifying the certain correlation threshold,
we detect the groups of nodes traversed by the essentially
correlated flows of random walkers. The structures of essentially
correlated flows and their appearance are the individual
characteristics of a city. We measure the quantity and extension
of clusters of nodes sharing the essentially correlated flows of
random walkers by an information parameter reflecting the
complexity of random walk traffic.

In Sec.~5, we discuss the statistical mechanics of so called {\it
lazy random walks} specified by the parameter $0< \beta \leq 1$.
In the model of lazy random walks, an agent located at a node $v$
moves to a neighboring node with probability   $\beta k_v^{-1}$,
but rests modeless with probability $(1-\beta)$. We compute the
well-known thermodynamical quantities (the internal energy,
entropy, the free energy,
 and pressure) describing the macroscopical states of an
ensemble of "lazy" random walkers on the dual graphs of compact
urban patterns. We provided the detailed interpretation for each
thermodynamical parameter in the context of random walks.

\section{Access times in the compact city structures}
 \noindent

The important characteristics of random walks defined on finite
graphs is the {\it access time} $H_{ij}$, the expected number of
steps before node $j$ is visited, starting from node $i$
\cite{Lovasz}. The elements of matrix $H_{ij}$ are computed for
each pair
 of nodes $i,j$
 following the formula:
\begin{equation}
\label{access}
H_{ij}=2m\sum_{s=2}^N\frac 1{1-\mu_s}\left(\frac{\varphi^2_{sj}}{k_j}-
\frac{\varphi_{si}\varphi_{sj}}{\sqrt{k_ik_j}}\right),
\end{equation}
in which $\mu_1=1>\mu_2\geq\ldots\mu_N\geq -1$ are  the eigenvalues and
$\varphi_{s}$ are the relevant
 eigenvectors of the symmetrized transition matrix $D^{-1/2}TD^{1/2}$.

The access time from $i$ to $j$ may be different from the access time from $j$
to $i$, $H_{ij}\ne H_{ji}$, even in a regular graph. A deeper symmetry property
of access times
for undirected graphs was discovered
in \cite{Copersmith},
\begin{equation}
\label{triangle}
H_{ij}+H_{jk}+H_{ki}=H_{ik}+H_{kj}+H_{ji}
\end{equation}
for every three nodes in $G$. This property allows for the ordering of nodes in
 the graph with respect to their accessibility for the random walkers.
It has been pointed out in \cite{Lovasz} that the nodes of any graph can be
ordered so that if $i $ precedes
$j$ then $H_{ij}\leq H_{ji}$. This ordering is not unique, but one can partition
the nodes by putting $i$
and $j$ in the same equivalence class if $H_{ij}=H_{ji}$ and obtain  a well-defined
ordering of the equivalence
classes, independent of any reference node. The nodes in the lowest class are
difficult to reach but easy to get out of, while the nodes in the highest class
 are easy to reach but
difficult to leave. If a graph has a vertex-transitive automorphism group, then
 $H_{ij}=H_{ji}$ for
all $i,j\in G$. The random target identity \cite{LovWinkl},
\begin{equation}
\label{target}
\sum_j\pi_jH_{ij}=\mathrm{Const},
\end{equation}
states that the expected number of steps (the random target access
time, $\tau$) required to reach a node randomly chosen from the
stationary distribution $\pi$ is a constant, independent of
 the starting point of the given graph $G$. The values of random target access
  times grow with the size of
graphs and are very sensitive to their structures. Their values
for the compact urban structures are given in the last column of
Tab.~1.

The properties of access times can be used in order to estimate the accessibility
of certain streets
and districts by random walkers starting from the rest of city. Computations of
access times
to the streets in the studied compact urban structures convinced us that for any
 given node
 $i$ the access times to it, $H_{ij},$ change with $j$ inferentially in comparison
 with their
typical values, and therefore, the mean access time,
\begin{equation}
\label{meanaccess}
h_i=\frac 1N\sum_{j=1}^n H_{ij}
\end{equation}
can be considered as a good  parameter for estimating the
accessibility of a street by random walkers. Distributions of mean
access times, $\alpha_h$, in  the city can be consider as
estimations of its connectedness. In particular, it helps to
detect the ghettos, the groups of streets almost isolated (in the
dynamical sense) from the rest of town. In Fig.~2, we have
presented together the distributions of mean access times to the
streets in Manhattan (dashed line) and Rothenburg o.d.T. (solid
line). The distribution for Rothenburg o.d.T. exhibits a local
maximum at relatively long access times ($\approx 300$ steps)
indicating that there is a number of low accessible streets in the
town. The distributions of mean access times to the canals in
Amsterdam (dashed line) and Venice (solid line) is presented on
Fig~3.

The distribution of  mean access times in the downtown of Bielefeld is of
essential interest since it comprises of
two structurally different parts (see Fig.~4.a).
The part "A" keeps its original structure (founded in
XIII-XIV cs.), while the part "B" had been subjected to the partial redevelopment in the
XIX-th century (Fig.~4.a). It is important to mention that
 the city districts constructed in accordance to different development principles and
in different historical epochs can be easily visualized on the
dual graph of the city. In Fig.~4.b, we have shown the
3D-representation of the dual graph of the Bielefeld downtown. The
$(x_i,y_i,z_i)$ coordinates of the $i-$th vertex of the dual graph
$G$ in three dimensional space are given by the relevant $i-$th
components of three eigenvectors $u^{(2)}$, $u^{(3)}$,and
$u^{(4)}$
 of the adjacency matrix $A_G$ (which does not coincide with the transition matrix $T$).
 These eigenvectors
 correspond  to the second, third, and fourth largest (in absolute value) eigenvalues
of $A_G$ \cite{Maple}. The 3D-dual graph of Bielefeld displays
clearly a structural difference between "A" and "B" parts: in 3D-
representation, the relevant subgraphs are located in the
orthogonal planes (Fig.~4.b). Sometimes other symmetries of dual
graphs can be discovered visually by using other triples of
eigenvectors if the number of nodes in the graph is not too large.

 In Fig.~5, we have displayed the distributions of mean access times $h$ to the
 streets located in the medieval
part "A" starting from those located in the same part of Bielefeld
downtown, from "A" to "A" (solid line). It has been computed by
averaging $H_{ij}$ over $i,j\in A$ in (\ref{meanaccess}). The
dashed line presents the distribution of mean access times to the
streets located in the modernized part "B" starting from the
medieval part "A" (from "A" to "B", $i\in A$ and $j\in B$). One
can see that in average in takes longer time  to reach the streets
located in "B" starting from "A". The similar behavior is
demonstrated by the random walkers starting from "B" (see Fig.~6):
in average, it requires longer time to leave a district for
another one. Study of random walks defined on the dual graphs
helps to detect the quasi-isolated districts of the city.

\section{Dynamical modularity in the compact city \\ structures}
 \noindent

Dynamical modularity is a particular division of the graph into
groups of nodes on which the certain modes of diffusion process
are localized.
 It can be
detected by analyzing spectral properties of the relevant dual
graph that is of the spectrum of some differential operator
defined on it \cite{Colin98}. In particular, for the undirected
graphs, it is convenient to consider differential elliptic
self-adjoint operators with a positive spectrum. We have studied
the Laplace operator defined by (\ref{01}).

The eigenvalues of Laplace operator (\ref{01}) along with the continuous approximations of
their densities for the dual graphs of compact urban patterns are shown in
Figs.~7-9. The similarity of spectra allows us to divide the studied compact urban
patterns into three categories: i) the medieval cities: Bielefeld (Fig.~7) and Rothenburg
o.d.T.; ii) the canal patterns: Venice (Fig.~8) and Amsterdam; iii) the spectra
 with a highly degenerated
eigenmode ($\lambda=1$): Manhattan (Fig.~9).

It is important to note that the densities of eigenvalues for the {\it compact} urban patterns
differ dramatically from those computed for the classical random
graphs of Erd\"{o}s-R\'{e}nyi model \cite{13,14}, deviate form the
semicircular law, and the densities found for the scale-free
random tree-like graphs in \cite{DGMS}.
 In \cite{ESMS}, the density of eigenvalues for the Internet graph on the Autonomous
Systems (AS) level had been computed. The Internet spectrum on the AS level is
broadly distributed with two symmetric maxima and similar to
the eigenvalue density of random scale-free networks. In
contrast to them, the
spectral density  for the compact city samples are
bell shaped and tend to turn into a sharp peak localized at
$\lambda=1$ due to the highly degenerate 1-mode which
 would score a valuable fraction of  eigenvalues ($48\%$ for Manhattan).
  1-modes come in part from
pairs of streets of minimal connectivities branching of either
"broadways" or belt roads. These structures are overrepresented in
the compact city patterns.

The slowest modes of diffusion process  (\ref{02}) allow one to
detect the city modules characterized by the individual
accessibility properties. Indeed, these random walks do not
describe the properties of an actual city traffic, since the
approach does not concern the lengthes of streets referring
uniquely to the topological properties. However, the modularity
plays an important role in any local dynamical process taking
place in the city including its traffic conditions. The primary
feature of the diffusion process in the compact urban patterns is
the flow between the dominant pair of city modules: the
"broadways" and the relatively isolated streets remote from the
primary roads.

Due to the proper normalization, the components of the
eigenvectors $\mathbf{n}^{(\alpha)}$  play the role of the
Participation Ratios (PR) which quantify the effective numbers of
nodes participating in a given eigenvector with a significant
weight. This characteristic has been used in \cite{ESMS} and by
other authors to describe the modularity of complex networks.
However, under the abundance of a highly degenerate mode PR is not
a well defined quantity (since the different vectors in the
eigensubspace corresponding to the degenerate mode would obviously
have different PR).

As time advances the distribution of random walkers approaches a
steady state $ n_i^\infty\propto {k_i},$ in which all diffusion
currents are balanced. It corresponds to the principal eigenvector
$n_i^{(1)}$ related to the smallest eigenvalue of $L$. The
relaxation processes toward the steady state are described by the
remaining eigenvectors ${\mathbf n}^{(\alpha)}$, $\alpha>1$, with
the characteristic decay times $\tau^{(\alpha)}$, such that
$\exp(-1/\tau^{(\alpha)})=\lambda_\alpha$. The second smallest
eigenvalue of the scaled Laplacian (\ref{01}) is related to the
graph diameter,
$
\mathrm{diam}(G) \leq - {\ln (N-1)}/{\ln (\lambda_2)},
$
the maximum distance between any two vertices in the graph. It is
also related to the Fiedler vector
\cite{F} describing the algebraic connectivity of the graph. Namely,
let us consider the components of the
 eigenvector ${\mathbf n}^{(2)}$ corresponding to the second smallest
eigenvalue $\lambda_2$ for the scaled
Laplacian (\ref{01}) defined on a connected graph $G(V,E)$. Define
$V_1=\{v\in V: n_v^{(2)}<0\}$ and $V_2=\{v\in V: n^{(2)}_v\geq
0\}$, then the subgraphs induced by $V_1$ and $V_2$ are connected.

In general, each nodal domain on which the components of the
eigenvector ${\mathbf n}^{(\alpha)}$ does not change sign refers
to a coherent flow (characterized by its decay time
$\tau^{(\alpha)}$) of random walkers toward the domain of
alternative sign. The nodal domains participate in the different
eigenmodes as one degree of freedom, and therefore their total
number is important for detecting the dynamical modularity of city
networks. It is known from \cite{DGLS} that the eigenvector
${\mathbf n}^{(\alpha)}$ can have at most $\alpha+m_\alpha-1$
strong nodal domains (the maximal connected induced subgraphs, on
which the components of eigenvectors have a definite sign) where
$m_\alpha$ is the multiplicity of the eigenvalue $\lambda_\alpha$,
but not less than 2 strong nodal domains ($\alpha>1$)
\cite{BHLPS}. However, the actual number of nodal domains can be
much smaller than the bound obtained in \cite{BHLPS}. In the case
of degenerate eigenvalues, the situation becomes even more
difficult because this number may vary considerably depending upon
which vector from the $m_\alpha$-dimensional eigenspace of
degenerate eigenvalue $\lambda_\alpha$ is chosen. A fragment of
nodal matrix for the Chelsea village (Fig.~10.a) in the Manhattan island is
shown on Fig.~10.b: the components of all eigenvectors localized on the
nodal domains displayed in white (black) have always the positive
(negative) sign.

We investigate the statistics of components of eigenvectors
$\mathbf{n}^{(\alpha)}$ localized on a given street,
$\{n^{(\alpha)}_i\}_{\alpha}.$  To uncover a fine modular structure
of compact city patterns, we have computed the linear correlation
coefficients between the lists of eigenvector components for all pairs of streets in a city,
\begin{equation}\label{corr}
  C_{ij}= \frac{\mathrm{Cov}\left(
\{n^{(\alpha)}_i\}_{\alpha},\{n^{(\alpha)}_j\}_{\alpha}\right)}
{\sqrt{\mathrm{Var}\left(\{n^{(\alpha)}_i\}_{\alpha}\right)
\mathrm{Var}\left(\{n^{(\alpha)}_i\}_{\alpha}\right)}}.
\end{equation}
The linear correlation measures how well a linear function
explains the relationship between two data sets. The correlation
is positive if an increase in the eigenvectors components
localized on a given street corresponds to an increase in those
related to other street, and negative when an increase in one
corresponds to a decrease in the other.

The study of correlations between the components of eigenvectors
allows us for a precise  recovering of all dynamical modules in a
city, street by street. By tuning the sensitivity threshold
$\kappa>0$, one can detect the groups of streets characterized by
significant pairwise correlations $C_{ij}\geq \kappa$. In case of
a degenerate eigenmode, the whole bunch of streets of minimal
accessibility joins a correlation cluster at once. Investigating
the size of clusters characterized by the significant pairwise
correlations between the eigenvectors components, we have found
that it is very sensitive to the threshold value $\kappa$. Pairs
of correlated streets can be detected for $\kappa\leq \kappa_c$,
below the critical value $\kappa_c$ individual for each city.  For
instance, the strongest correlations in the Manhattan island
($\kappa \geq 0.16$) are observed between the Bowery st.
(Chinatown)  and the 6-th Avenue (Greenwich Village), and between
the Lafayette st. and Mott st. (Chinatown). By reducing  $\kappa$
just by  1$\%$, we immediately get many new correlated pairs and
[South FDR Dr., Allen st.], [Greene st.(Soho), Ave. D],
 [Crosby st. (Little Italy), Ave. D], [Centre st, Elizabeth st.], [Allen st., 6-th Ave.],
[Lafayette st., Ave. B], [Broadway, West st.] among the others.
In most cases  the streets in correlated pairs have the same
driving directions although the driving direction data has not been initially used in
the adjacency matrices. Reducing the value of sensitivity threshold, one can detect
the triples of correlated streets and more structurally complicated dynamical
modules. Then the correlated clusters merge exhibiting a sharp phase transition
into a giant connected component for essentially small $\kappa$.

The appearance of correlated $k$-tuples and their total numbers at
given $\kappa$ are by no means random and encode the important
{\it information} on the city connectedness. The complexity of
dynamical modularity can be measured by a quantity of information
encoded by the number of various $k$-tuples of essentially correlated nodes
at the different values of correlation threshold $\kappa$ in the following
way. Since the probability that a connected correlated $k$-tuple
appears in a {\it random} labelled graph of $N$ nodes is
\[
p_k(N)=\left(\begin{array}{c}
  N \\
  k
\end{array}\right)^{-1},
\]
its appearance corresponds to some quantity of information
$I=-p_k(N)\log_2 p_k(N).$ For a mesh of values $\kappa>0$, we had
counted the total numbers of all correlated $k$-tuples appeared in
each city,
 ${\mathcal N}_k(\kappa)$, and then found the
total amounts of information ${\mathcal I}(\kappa)$ encoded by the dynamical modularity
 in each studied
city pattern,
\begin{equation}
\label{Inf}
{\mathcal I}(\kappa)=-\sum_{k-{\mathrm tuples}}{\mathcal N}_kp_k\left(|V|\right)
\log_2p_k\left(|V|\right).
\end{equation}
It is worth to mention that in the absence of correlations as well as in the case when
 the {\it all} fluxes
of random  walkers though  nodes of the dual city graph are correlated, no information can be
encoded, and therefore  ${\mathcal I}=0.$ The results on the comparative information analysis
of dynamical modularity of the compact urban structures are displayed in Figs.~11-13.

In general, the quantity of information encoded by the dynamical
modularity of ancient cities exceeds that in the modern or
redeveloped ones since a number of correlated clusters of diverse
sizes could appear in the graphs with a less regular structure.
Information (via the threshold of correlations, $\kappa$)
calculated for Rothenburg and Bielefeld (subjected to a partial
redevelopment) are almost equal (Fig.~11). Both medieval cities
contain a number of streets characterized by the relatively highly
correlated traffic (they are the belt roads encircling the cities
along their fortress walls), while the traffic along the streets
close to the city centers appears to be less correlated, although
{\it all} streets of low accessibility join the correlation
cluster at once at some level $\kappa$, then the information on
the dynamical modularity turns to zero.

Information profiles obtained for the Venice and Amsterdam canal
networks look similar (see Fig.~12). However, it seems that
information in a message on possible correlations of gondoliers
traffic in Venice could be much more valuable than a cruise
schedule along the Amstel embankments. The triggering between
different information states displayed on Fig.~12 corresponds to
the merging of diverse correlated clusters into the bigger
correlated modules. The information peaks located close to
$\kappa=0$ indicate a phase transition to a giant correlated
component which cover  most of the nodes in the networks.
Nevertheless, {\it not all} canals join the giant correlated
module (since the amount of information is not zero as $\kappa\to
0$, but, on the contrary, turns to be high). It happens because of
the presence of an {\it anti}-correlated module in the canal
network, in which the components of Laplacian eigenvectors are
negatively correlated, $C_{ij}<0$.

The scale of the graph shown in Fig.~13 is incompatible with those
of Fig.~11 and Fig.~12 since the dynamical modularity in the
Manhattan island (one of the most regular city grids in the world)
contains essentially less information (it comes primarily from the
correlated clusters risen in the region of Central park) than that
in the ancient cities.

\section{Statistical mechanics of lazy random walks in the compact city structures}
 \noindent

A number of different "measures" quantifying the various
properties of complex networks has been proposed in so far in a
wide range of studies in order to distinguish the groups of nodes
and enlighten the relations between them. Some measures can be
computed directly from the graph adjacency matrix: {\it
likelihood} \cite{LAWD}, {\it assortativity} \cite{assort}, {\it
clustering} \cite{clust}, {\it degree centrality} \cite{WF},
\cite{d_cent}, \cite{Free}, {\it betweeness centrality}
\cite{Free}, {\it link value} \cite{link_v}, {\it structural
similarity} \cite{LHN}, {\it distance} (counting the number of
paths between vertices) \cite{distance}. Other measures (concerned
with the networks embedded into Euclidean space) involve the
lengths of links or the true Euclidean distances between nodes:
{\it closeness centrality} \cite{WF}, \cite{clos_c}, {\it
straightness centrality} \cite{straight_c}, \cite{VM_PRL},
 {\it expansion} \cite{link_v}, {\it information centrality} and {\it graph efficiency}
\cite{VM_PRL},\cite{efficiency}. A good summary on the several
centrality measures can be found in \cite{chaos} and in \cite{dK}
for the Internet related measures.  The list of available measures
is still far from being complete, the new measures appearing
together with any forthcoming network model. It is also worth to
add some spectral measures (concerned the eigenvalues of graph
adjacency matrix): {\it subgraph centralization} \cite{CRS}, {\it
subgraph centrality} \cite{ERV05}, {\it network bipartivity}
\cite{ER_PRE} and many others.

In the present section, we study the statistics of flows of lazy
random walkers roaming in the compact city patterns. The random
walkers have no mass and do not interact with each other, so that
they do not neither to contribute  to the energy nor to the
momentum transfer. Nevertheless, their flows have the nontrivial
thermodynamical properties induced by the complex topology of
streets and canals they flow along. The obvious advantage of
statistical mechanics
 is that the mathematical objects we introduce and all relations between various
statistical quantities are well known in the framework of
thermodynamic formalism.

In the following, we use the inverse temperature parameter
$\beta>0$ which can be considered either as an effective time
scale in the problem (the number of streets a walker passes in one
time step) or as the laziness parameter defined in Sec.~2. Since
the eigenvalues $\lambda_\alpha$ of scaled Laplacian operator
(\ref{02}) are positive and bounded, one can define  for them
three well known spectral functions:
\begin{enumerate}
\item
the heat kernel (the partition function),
\begin{equation}
\label{K}
K(\beta)=\mathrm{Tr} \exp \left(-\beta{L}\right)= \sum_{\alpha=1}^{N} \mathrm{e}^{-\beta\lambda_\alpha},
\end{equation}
converging as $\beta\geq 0$, for $N\to \infty,$
\item the spectral zeta function (the spectral moments)
\begin{equation}
\label{zeta}
\zeta(s)=\sum_{\alpha=1}^{N} \lambda_\alpha^{-s},
\end{equation}
\item and the spectral density,
\begin{equation}
\label{rho}
\rho(\lambda)=\sum_{\alpha=1}^N\delta(\lambda-\lambda_\alpha).
\end{equation}
\end{enumerate}
These spectral functions are related to each other by the Laplace transformation,
\begin{equation}
\label{K2rho}
K(\beta)=\int_0^{\infty}d\lambda{\ } {\mathrm e}^{-\beta\lambda} \rho(\lambda),
\end{equation}
and by the  Mellin transformation (up to the $\Gamma$-function) \cite{TE},
\begin{equation}
\label{zeta2K}
\zeta(s)=\frac 1{\Gamma(s)}\int_0^{\infty} \beta^{s-1}K(\beta )d\beta.
\end{equation}
Furthermore,
\begin{equation}
\label{zeta2rho}
\zeta(s)=\int_{-\infty}^{\infty} \rho(\lambda)\lambda^{-s}d\lambda.
\end{equation}
Let us note that since $0<\lambda_\alpha<2$, we immediately obtain
 bounds for the spectral moments,
\[
\zeta(-n)<N\cdot 2^{n}, \quad \zeta(n) > N\cdot 2^{-n}.
\]
The partition function $K(\beta)$ meets the conditions of
Bernstein's theorem \cite{Feller}:
\[
(-1)^nK^{(n)}(\beta)\geq 0, \beta>0, n\geq 1,
\]
and therefore defines a unique non-negative measure
$d\mu(\beta)=K(\beta)d\beta$ on $\mathbb{R}_+$ providing the
solutions for the Hamburger and Stiltjes moments problems,
\cite{TE}. Concerning the list of "measures" given above,  it is
worth to note that, in principle, any of spectral functions
(\ref{K}-\ref{rho}) at a given temperature $\beta^{-1}$ can be
used as a measure of some quantities relevant to a certain lazy
random walk model.

The ensemble of random walkers on the graph $G(V,E) $ can be
characterized by the following {\it macroscopical} quantities: 1)
the internal (averaged) energy, $\bar E=-\partial_\beta\ln
K(\beta)$; 2) the entropy, $S=\ln K(\beta)+\beta \bar E$; 3) the
free energy, $F=\beta^{-1}\ln K(\beta)$, and 4) the pressure,
$P=\beta^{-1}\partial_{|V|}\ln K(\beta).$ In the last equality, we
have considered the differential with respect to the graph size as
$d|V|=\rho(\lambda)d\lambda$. The thermodynamics of compact urban
city patterns is presented on the Figs.~14-22 and in Tab.~2. The
collected data give an insight into the various aspects of {\it
averaged} streets (canals) accessibility in a given city.

Due to the complicated topology of streets and canals, the flows
of random walkers exhibit the spectral properties similar to that
of a thermodynamical system characterized by a nontrivial internal
energy (Figs.~14-16). It grows with temperature (the inverse
parameter of lazy random walks)
 $\beta^{-1}$ (albeit still bounded in the interval
$[0,2]$). As usual, the absolute value of internal energy relevant
to a given city cannot be precisely measured, but we can measure
its difference in any temperature interval.
In principle, the slopes of internal energy curves are steeper (the internal energy
grows faster
with $\beta^{-1}$) in the modern cities with a quite regular grid-like structures, in which
any street has a relatively good accessibility.

In thermodynamics, entropy is an extensive state function that
accounts for the effects of irreversibility in thermodynamic systems.
It describes the number
 of the possible microscopic configurations of the system.
In the problem of finite random walks, its value quantifies the
diversity of flows which can be detected in the city at a given
temperature and grows with temperature. In Figs.~17-19, we have
displayed the entropy curves via $\beta^{-1}$. It is important to
mention that the improve of street accessibility causes a decrease
of entropy growth rates for large temperatures (small $\beta$).
The entropy of random walker flows as well as its growth rate in
Manhattan is ever less than in any other compact city structure we studied.

Due to the numerous junctions and the highly
entangled meshes of city street and canals,  the random walkers loss their ways
and it takes long time for them to cross a city roaming randomly along the streets.
This can be interpreted as an effect of a slight negative pressure
involving the random walkers into the city. It is obvious that the strength of
pressure should vary from one district to another within a city and can be different
 for the different modes of diffusion process which have no rigorous bind to the city
 administrative units being localized on the certain groups of streets and canals.
We have computed the pressure spectra $P(\lambda)$  forcing the
flows of random walkers with eigenmodes $\lambda$ into the compact
city structures (see Figs.~20-22). Generally speaking, the more
junctions a city has, the stronger is the drag force: despite
Venice has more canals than in Amsterdam, the number of junctions
between them is less than in Amsterdam, and therefore the negative
pressure in the latter city is stronger (Fig.~20). The numbers of
streets in Rothenburg and in the downtown of Bielefeld are equal,
but there are more crossroads in Bielefeld than in Rothenburg
(Fig.~21).

One can see that pressure profiles have maxima close to
$\lambda=1$ (correspondent to a minimal drug force) for Amsterdam,
Venice, and Manhattan. It calls for the idea of a "transparency
corridor", i.e. a sequence of streets and junctions (on which the
relevant eigenmodes are localized) along  which the city can be
crossed at minimal time. In Manhattan, the pressure profile is
almost zero at $\lambda=1$. The multiply degenerated eigenmode
$\lambda=1$ appears in Manhattan due to numerous junctions of low
accessible streets to Broadway and FDR Drive. In Venice, the
minimal pressure is achieved on the Grand Canal and Giudecca
Canal, and it is due to Het IJ and Amstel river in Amsterdam.

In Tab.~2,we have sketched the figures for the internal
energy, entropy, the free energy, and heat capacity for all studied
cities at $\beta=1$ (for the usual random walks
model), in purpose of comparison. We also
 gave the location of maxima for the heat capacity profiles.

 \section{Discussion and Conclusion}
 \noindent

We have studied the finite Markov chain processes defined on the
dual graphs of compact urban patterns. The traffic of random
walkers, indeed, do not describe the actual traffic conditions in
the city patterns that we have analyzed, but concerns their
topological properties giving a sense to the notions of
accessibility and modularity. Our approach can be readily used in
order to investigate connectedness and efficiency of
transportation lines in different complex networks.

The methods developed in graph theory and in probability theory
give us a detailed picture of local and global properties of city
structures. Dynamical modularity representing the localization of
dynamical modes on certain streets and canals cannot be detected
neither from the adjacency matrix of a graph nor from the
transition probability matrix. We have studied the random walks of
a large number of massless particles introduced in the dual graphs
of compact cities and investigated their properties by means of
statistical mechanics. The spectrum of Laplace operator is broadly
distributed for the compact city patterns and has a bell shape, in
general.  However, it turns into a sharp peak if there is a number
of low accessible streets branching of a prime street in a city.

To detect the dynamical modularity precisely, we have computed the
linear correlation coefficients between the lists of eigenvector
components for all pairs of streets (canals) in the cities.
Tuning the sensitivity threshold of correlations, one can detect
the various modules of essentially correlated streets.  The
complexity of dynamical modularity can be measured by the
information quantity. The information versus correlation profile
is an individual city fingerprint.

The statistical mechanics description of random walks in the
compact cities provides us with a rigorous definition of the
unique measure and all statistical moments which have a definite
relation to the entire topological properties of complex networks.
Due to a complicated topology of city streets and canals, the
flows of random walkers acquire  nontrivial thermodynamical
characteristics which can be implemented in studies of
interactions between different city modules and of city stability
with respect to occasional accessibility problems that would be of
vital importance for the city traffic conditions in urgency.

The obvious advantage of the method is that the information on the role of a given street
or canal in the entire city network can be interpreted by the very end, and all city modules
can be named street by street.

The method can be generalized for the complex networks which can
be described by  directed or weighted graphs. For instance, one
can consider a graph  of city plan in which the weights of edges
are the actual lengthes of streets. In the case of weighted
adjacency matrix, the spectrum of the relevant Laplace operator
acquires the pairs of complex conjugated eigenvalues. Then the
components of eigenvectors are also complex, and the coefficients
of linear correlations should be computed for the real and
imaginary parts of spectrum separately. Other approach could be
used while studying  directed graphs for which the adjacency
matrix is not symmetric. For the random walks defined on the
directed graphs,
 the probability that a random walker enters a node
is not equal to the probability it leaves the node. The ensembles
of random walkers introduced on the directed graphs can be
described by  Nelson stochastic mechanics \cite{Nelson},
\cite{BlanchCombe}. The stationary configurations of random
walkers in such models can be studied by means of biorthogonal
decomposition \cite{Lima}.

\section{Acknowledgements}
 \noindent

One of the authors (D.V.)  is grateful to the Alexander von
Humboldt Foundation (Germany), to the Bielefeld-Bonn Stochastic
Research Center (BiBoS, Bielefeld, Germany), to
the DFG-International Graduate School "Stochastic and real-world
problems", to the Centre de Physique Theorique (CPT, C.N.R.S.
 Contract  CEA - EURATOM N006385)   for their financial
support and hospitality during the preparation of the present paper.

We thank R. Lima, B. Cessac, and G. Zaslavsky for  fruitful
discussions and multiple comments on our work.

\newpage

{\bf Table 1: Some features of dual graphs of the studied city patterns}

\vspace{1cm}
\begin{tabular}{||c||c|c|c|c||}
   \hline \hline
     & $|V|$ & $|E|$ &  diam($G$)& $\tau$
    \\ \hline\hline
 Rothenburg ob.d.T. & 50 & 115 &  13& 545 \\
Bielefeld downtown& 50 & 142 &  14 & 551 \\ \hline
 Amsterdam canals & 57 & 200 & 11 & 849\\
 Venice canals & 96 & 196 &  14 & 1550\\ \hline
  Manhattan & 355 & 3543 &  17 & 4557 \\
  \hline \hline
\end{tabular}

\vspace{0.5cm} {\bf Caption for Table 1:} Some features of dual
graphs of compact city patterns: $|V|$ is the number of streets,
$|E|$ the number of junctions (crossroads), the graph diameter
$\mathrm{diam}(G)$ is the maximal graph-theoretical distance
between any two vertices of the dual graph. In the last column,
the the random target access time $\tau$, the expected number of
steps required to hit a node randomly chosen in the city from the
stationary distribution $\pi$. The random target access time is
independent of the starting point.

\newpage

{\bf Table 2:  Thermodynamical properties of ensembles of random walkers in the compact
cities at $\beta=1$ (the usual random walks).}

\vspace{1cm}
\begin{flushleft}
\begin{tabular}{||c||c|c|c|c|c||}
  \hline \hline
     & ${\bar E}(\beta=1)$ & $S(\beta=1)$ & $F(\beta=1)$ &
     $C_{\mathrm{max}}$  & $\beta_{C_{\mathrm{max}}}$
    \\ \hline\hline
Rothenburg ob.d.T. & 0.946604  & 3.885109 &2.93850 &1.60468 & 11.1 \\
Bielefeld downtown&0.957324 & 3.89042 &   2.93309& 2.04195 & 9.90 \\ \hline
  Venice canals & 0.95082 & 4.53962 & 3.58880 & 1.92003 & 27.4 \\
 Amsterdam canals& 0.898409 & 3.88836 & 2.98996 & 1.3353; 1.1094 &  8.6; 48.75\\ \hline
  Manhattan & 0.990083 &5.86713 & 4.87705  & 5.87368  & 13.65 \\
  \hline \hline
\end{tabular}
\end{flushleft}

\vspace{0.5cm}
{\bf Caption for Table 2:}
$\bar E$ is the internal energy, $S$ is the entropy, $F$ is the free energy, $C_{\mathrm max}$
is the maximal values of heat capacity, and $\beta_{C_{\mathrm max}}$ is the
points they are achieved.

\newpage

\begin{figure}[ht]
 \noindent
\begin{center}
\epsfig{file=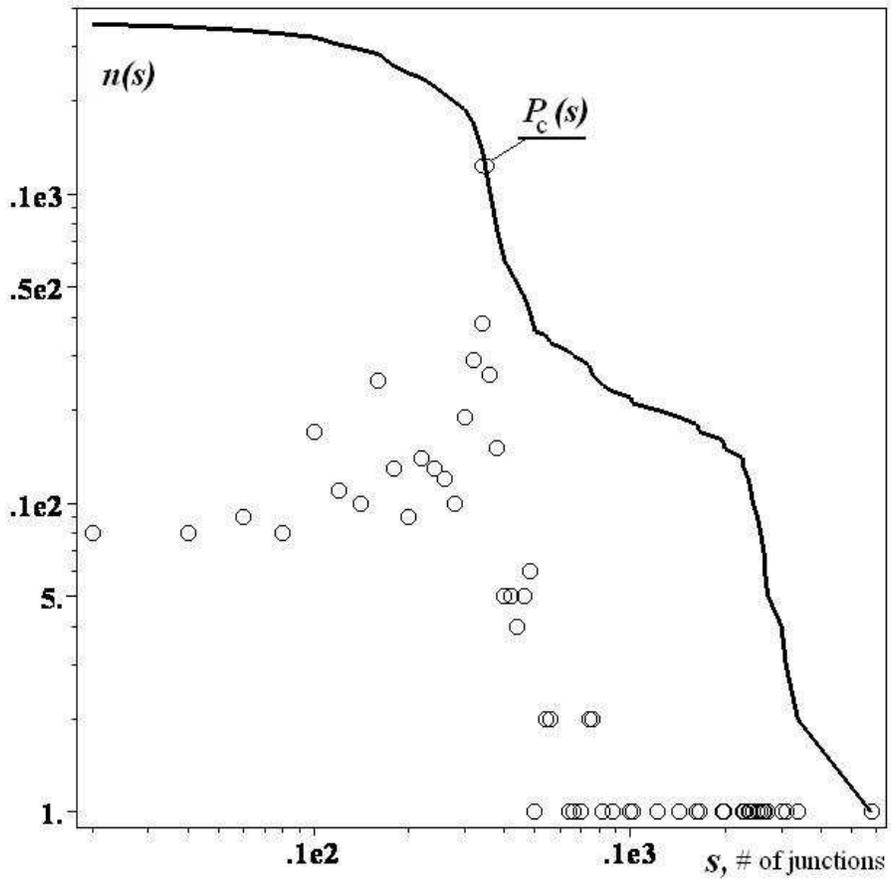, angle= 0,width=12.0cm,height=12cm}
\end{center}
\caption{The logarithm of fractions of streets $\ln n(s)$ via the
logarithm of number of their junctions $\ln s$ in Manhattan (shown
by points). The solid line represents for the logarithm of the
relevant cumulative distribution $\ln P_c(s)$ \cite{NewmanSIAM}.}
\end{figure}

\newpage

\begin{figure}[ht]
 \noindent
\begin{center}
\epsfig{file=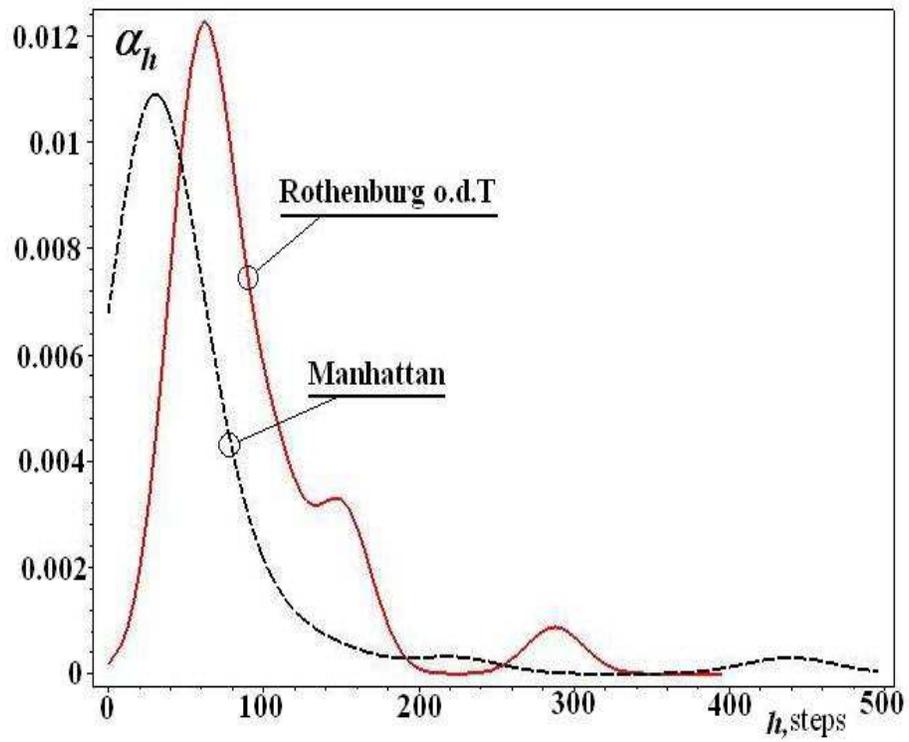, angle= 0,width=12.0cm,height=10cm}
\end{center}
\caption{The distributions of mean access times $h$ to the streets
in Manhattan (dashed line) and Rothenburg ob der Tauber (solid
line). The distribution $\alpha_h$ for Rothenburg o.d.T. exhibits
a local maximum at relatively long access times (~300 steps)
indicating the presence of a number of low accessible streets  in
the town.}
\end{figure}

\newpage

\begin{figure}[ht]
 \noindent
\begin{center}
\epsfig{file=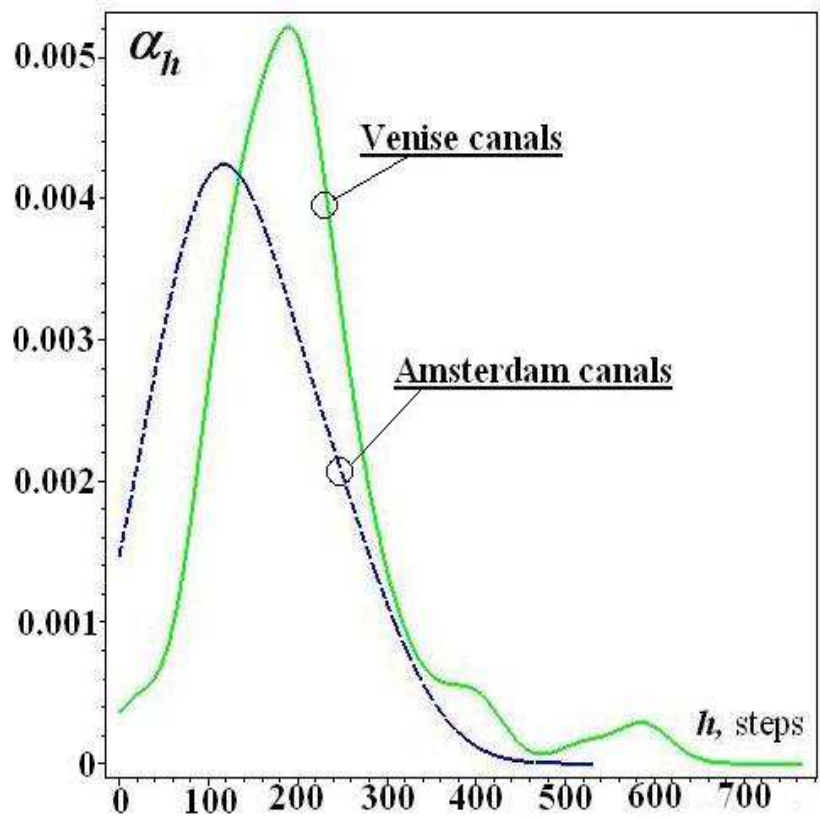, angle= 0,width=11.0cm,height=11cm}
\end{center}
\caption{The distributions of mean access times $h$ to the canals in Amsterdam
 (dashed line) and Venice (solid line).}
\end{figure}

\newpage
\thispagestyle{empty}

\begin{figure}[ht]
 \noindent
\begin{center}
\begin{tabular}{lr}
a.&\epsfig{file=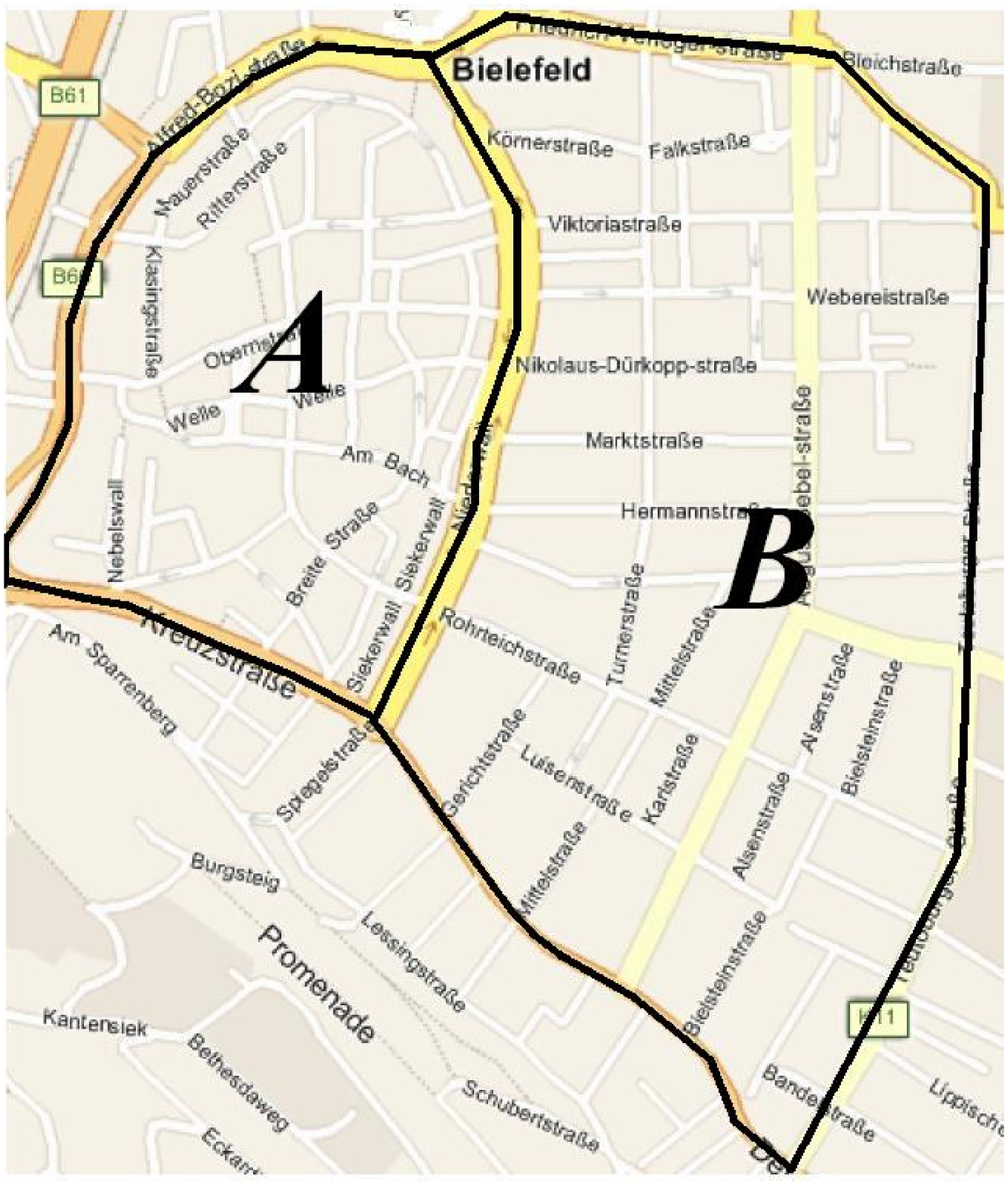, angle= 0,width=7.0cm,height=7cm}\\
b. &\epsfig{file=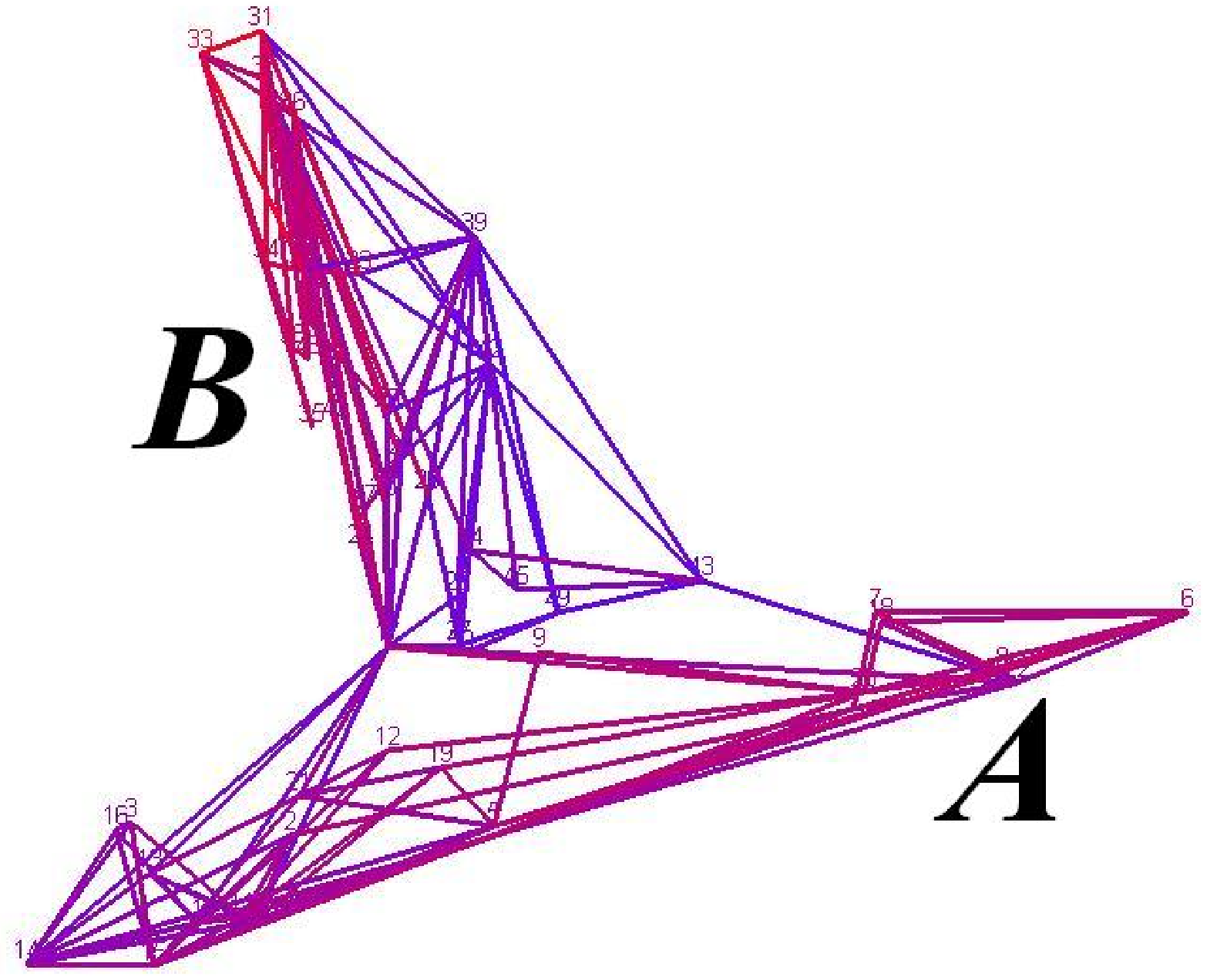, angle= 0,width=7.0cm,height=7cm}
 \end{tabular}
\end{center}
\caption{The city map of  Bielefeld downtown (a) is presented
together with its 3D representations of the dual graph (b). The
"A"-part  keeps its original structure (founded in XIII-XIV cs.);
the part "B" which had been redeveloped in the XIX-th century. The
$(x_i,y_i,z_i)$ coordinates of the $i-$th vertex of the dual graph
in three dimensional space are given by the relevant  $i-$th
components of three eigenvectors
 $u^{(2)}$, $u^{(3)}$,and $u^{(4)}$
 of the adjacency matrix $A$ of the dual graph }
\end{figure}

\newpage

\begin{figure}[ht]
 \noindent
\begin{center}
\epsfig{file=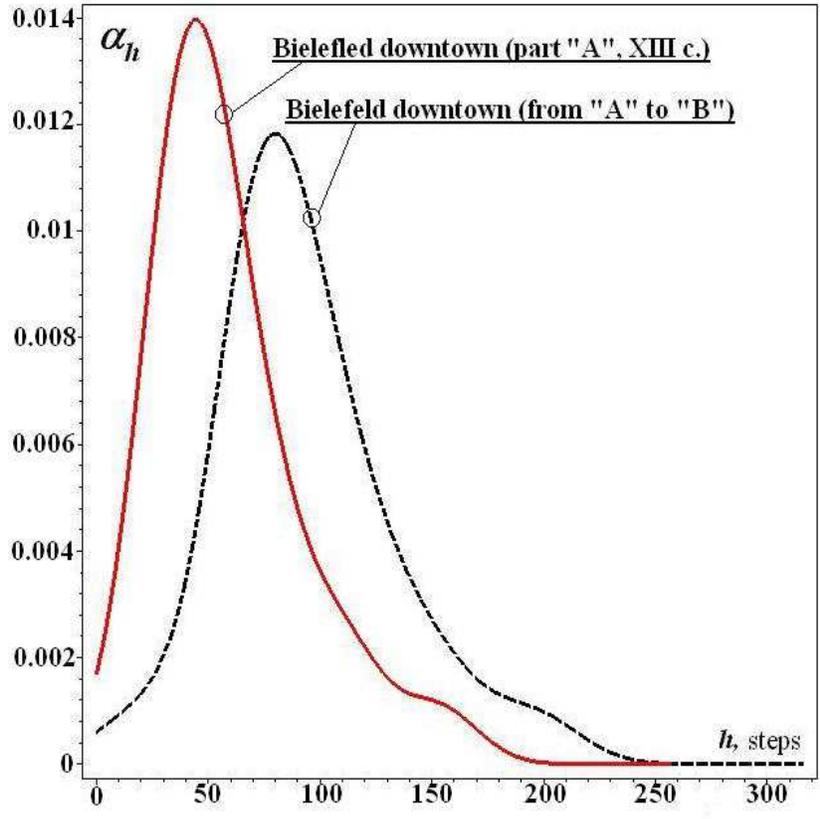, angle= 0,width=11.0cm,height=11cm}
\end{center}
\caption{The distributions of mean access times $h$ to the streets
located in the medieval part "A" starting from those located in
the same part of Bielefeld downtown, from "A" to "A" (solid line).
The dashed line presents the distribution of mean access times to
the street located in the modernized part "B" starting from the
medieval part "A" (from "A" to "B"). In average, in takes longer
time to reach the streets located in "B" starting from "A".}
\end{figure}

\newpage

\begin{figure}[ht]
 \noindent
\begin{center}
\epsfig{file=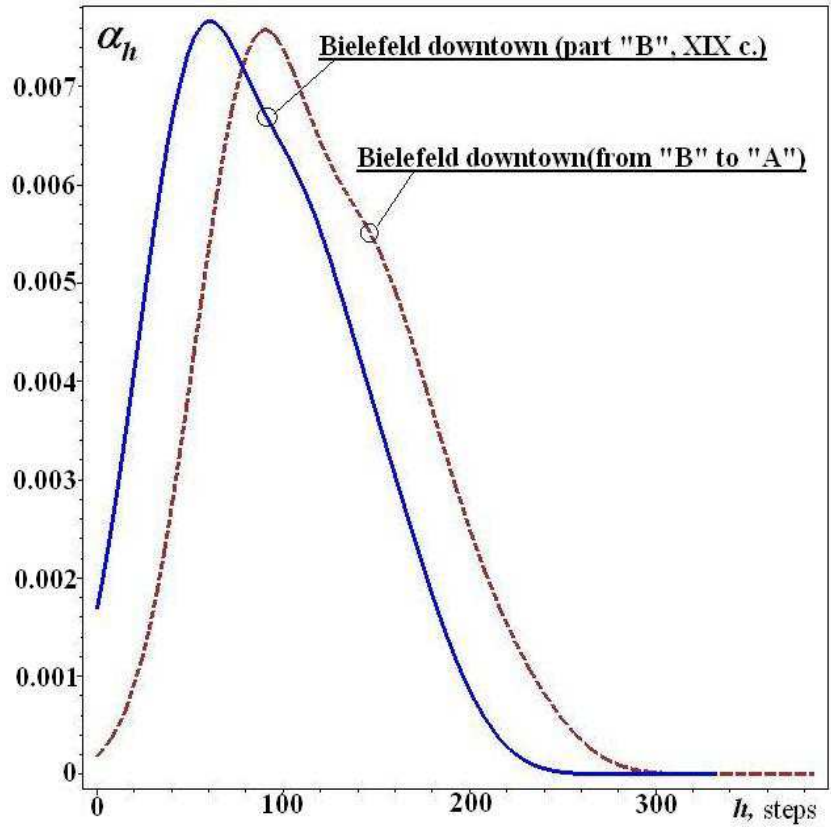, angle= 0,width=11.0cm,height=11cm}
\end{center}
\caption{The distributions of mean access times $h$ to the streets located in the "B"
part starting from "B" (solid line).
The dashed line presents the distribution of mean access times to the street
 located in the "A" part
starting from "B".}
\end{figure}

\newpage

\begin{figure}[ht]
 \noindent
\begin{center}
\epsfig{file=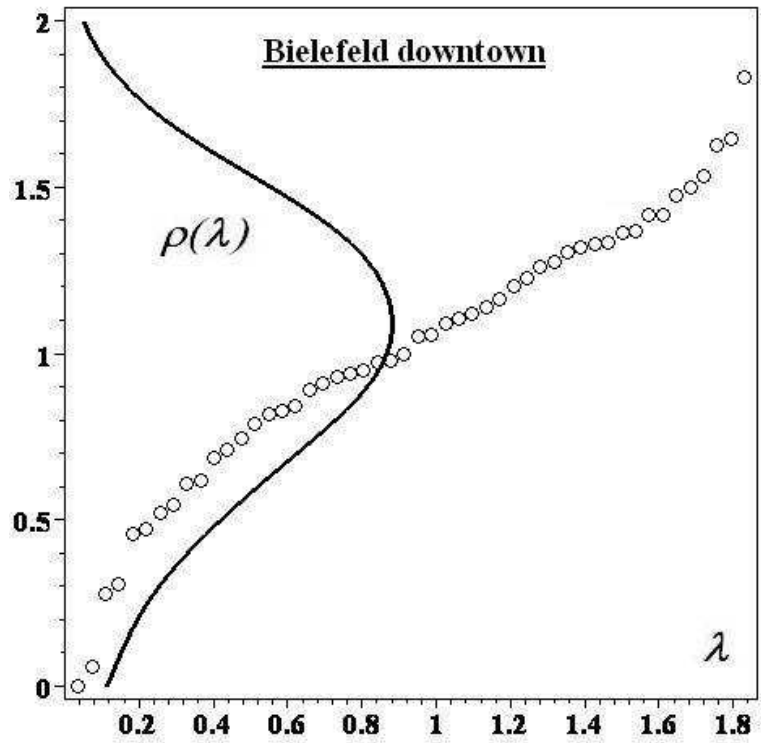, angle= 0,width=10.0cm,height=10cm}
\end{center}
\caption{The eigenvalues of the Laplacian  (\ref{01}) defined on
the dual graph of Bielefeld together with the continuous
approximation of its density.}
\end{figure}

\newpage

\begin{figure}[ht]
 \noindent
\begin{center}
\epsfig{file=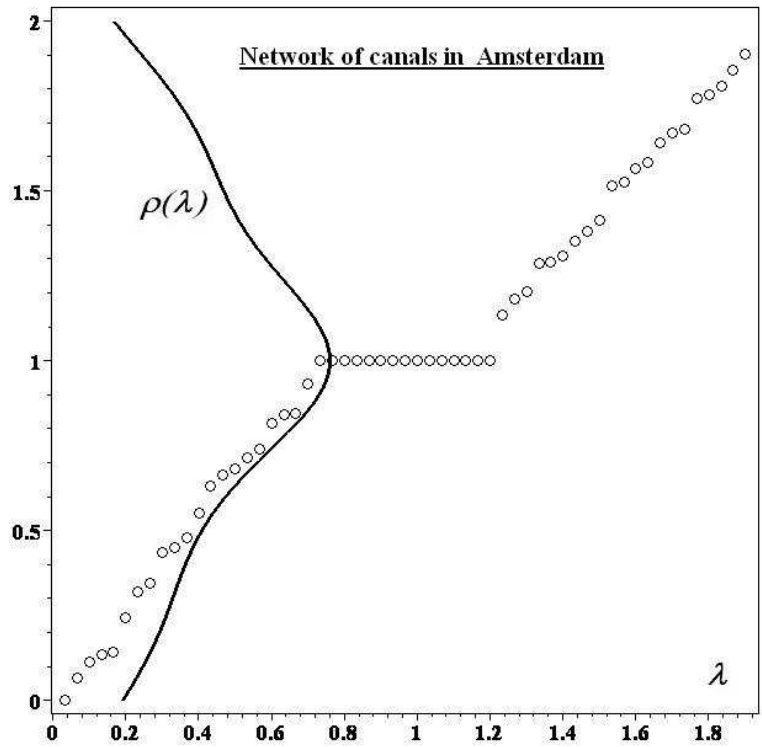, angle= 0,width=10.0cm,height=10cm}
\end{center}
\caption{The eigenvalues of the Laplacian  (\ref{01}) defined on
the dual graph of Amsterdam canal network together with the
continuous approximation of its density.}
\end{figure}

\newpage

\begin{figure}[ht]
 \noindent
\begin{center}
\epsfig{file=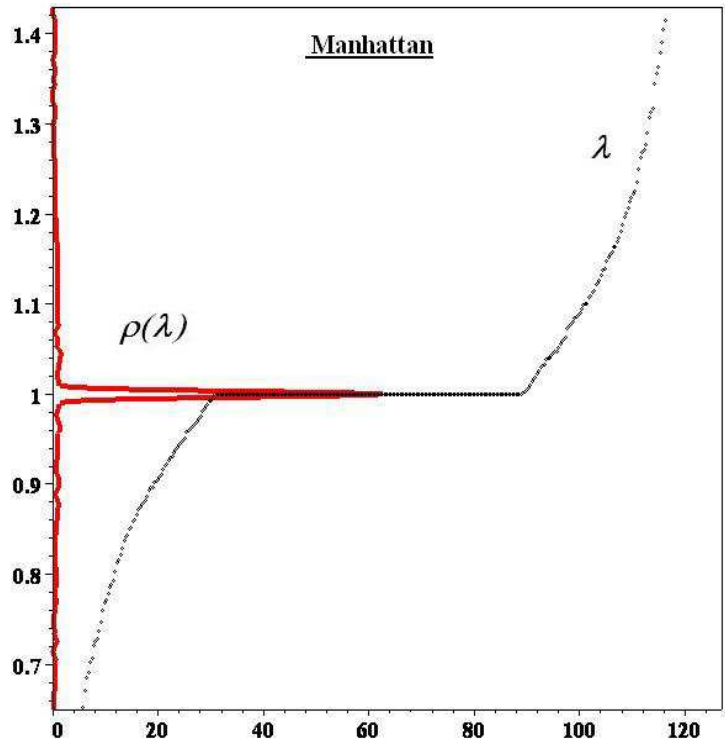, angle= 0,width=10.0cm,height=10cm}
\end{center}
\caption{The eigenvalues of the Laplacian (\ref{01}) defined on
the dual graph of Manhattan together with the continuous
approximation of its density.}
\end{figure}

\newpage

\begin{figure}[ht]
\noindent
  \begin{center}
\begin{tabular}{lr}
 a.&\epsfig{file=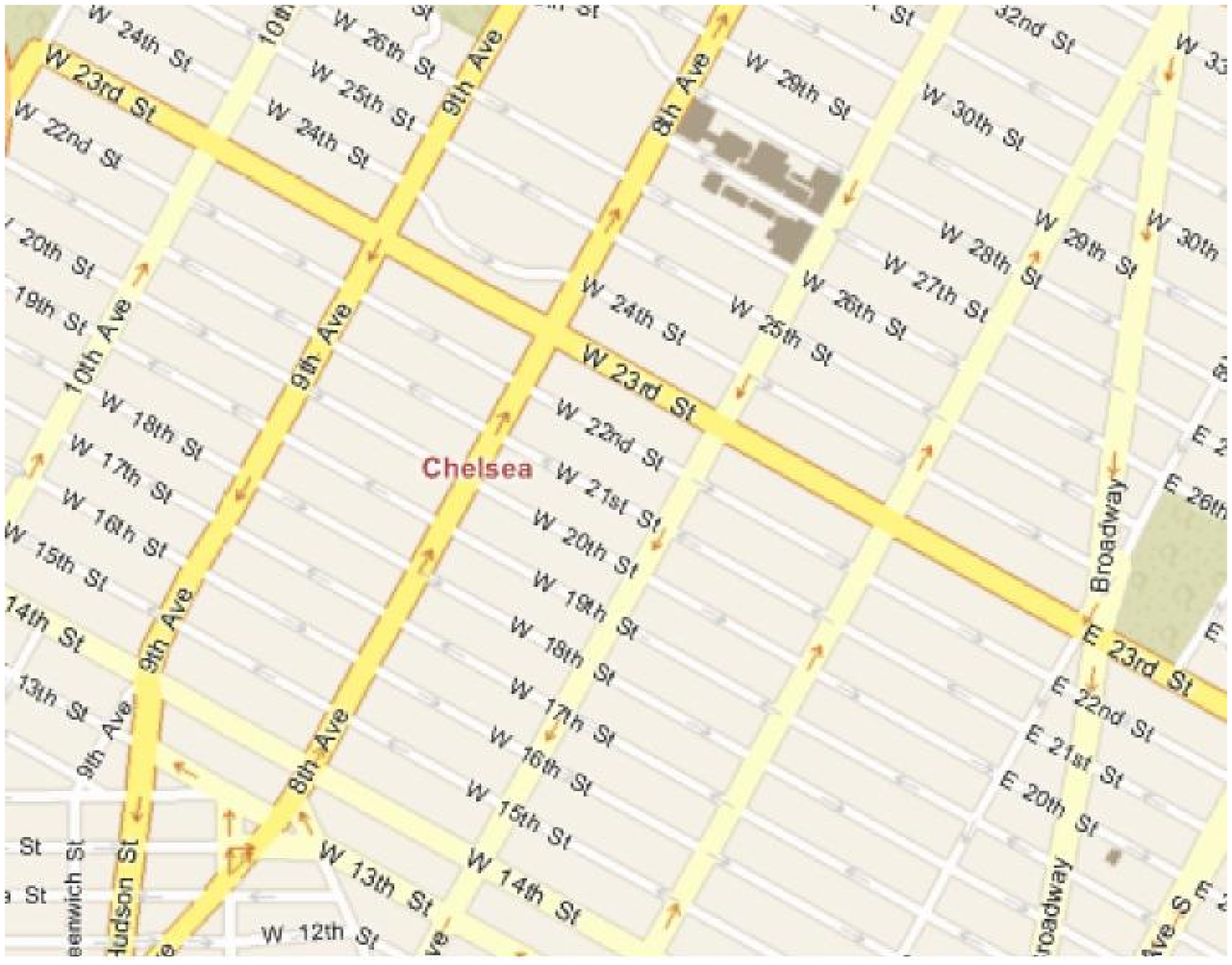, angle= 0,width=7cm,height=6cm}\\
b.&\epsfig{file=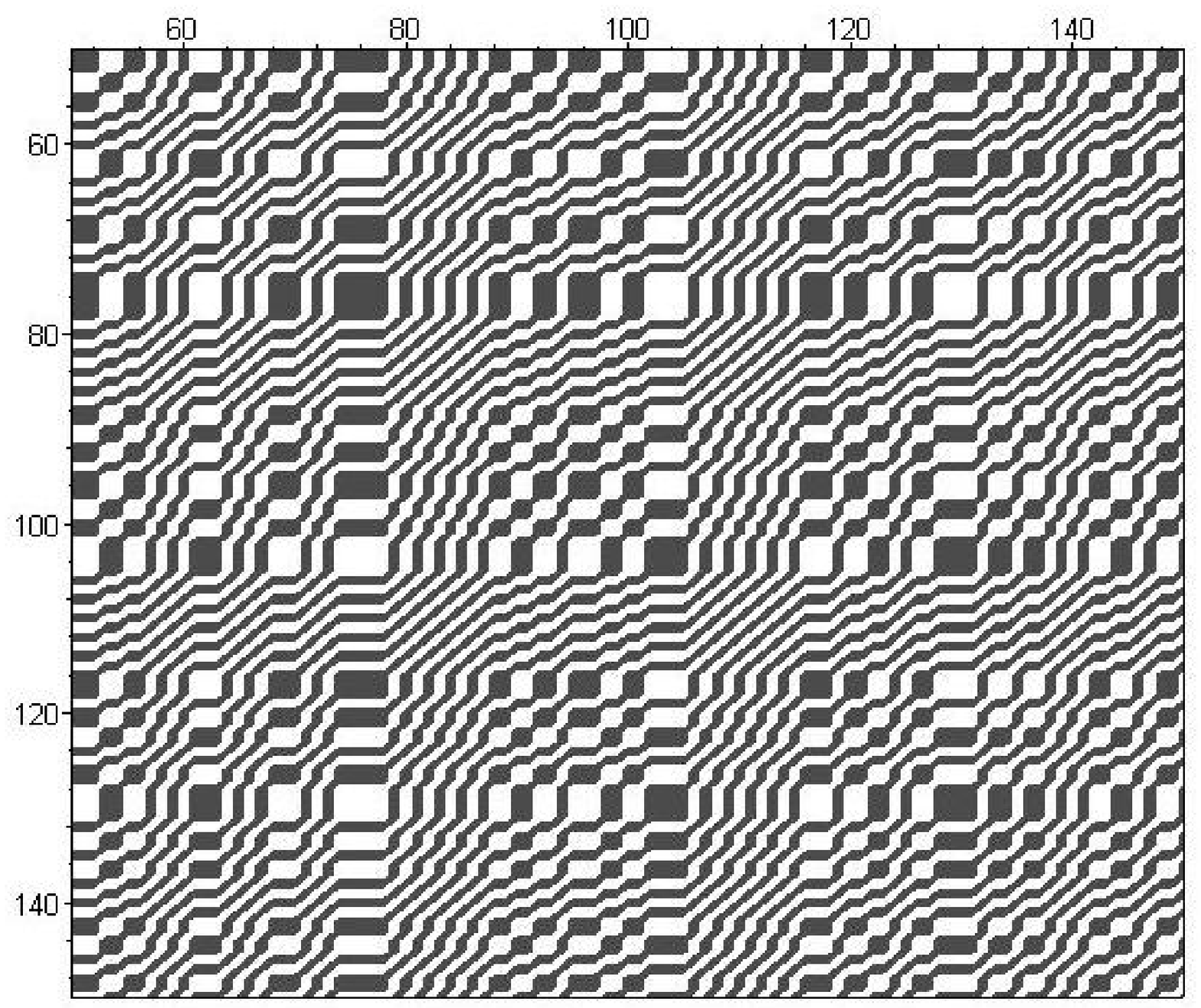, angle= 0,width=7.2cm,height=7cm}\\
 \end{tabular}
 \end{center}
\caption{a. Plan of the Chelsea village in Manhattan and the matrix
plot of its nodal domains (b).
All eigenvectors localized on the nodal domains displayed in white
(black) have always the positive
(negative) sign. }
\end{figure}

\newpage

\begin{figure}[ht]
 \noindent
\begin{center}
\epsfig{file=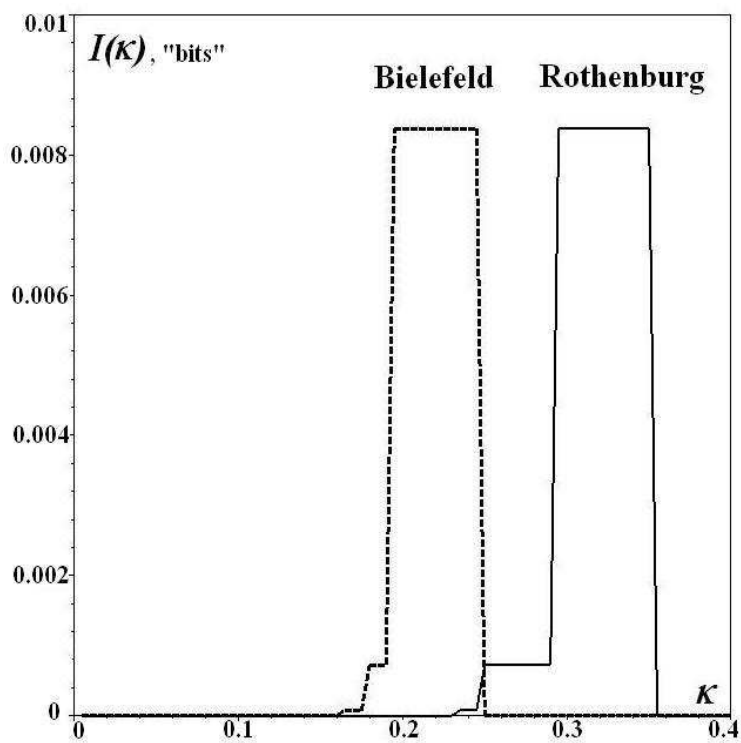, angle= 0,width=10.0cm,height=10cm}
\end{center}
\caption{Quantity of information (bits) encoded by the dynamical
modularities of Rothenburg and Bielefeld via the threshold of
essential correlations, $0\leq\kappa\leq 1$.  }
\end{figure}

\newpage

\begin{figure}[ht]
 \noindent
\begin{center}
\epsfig{file=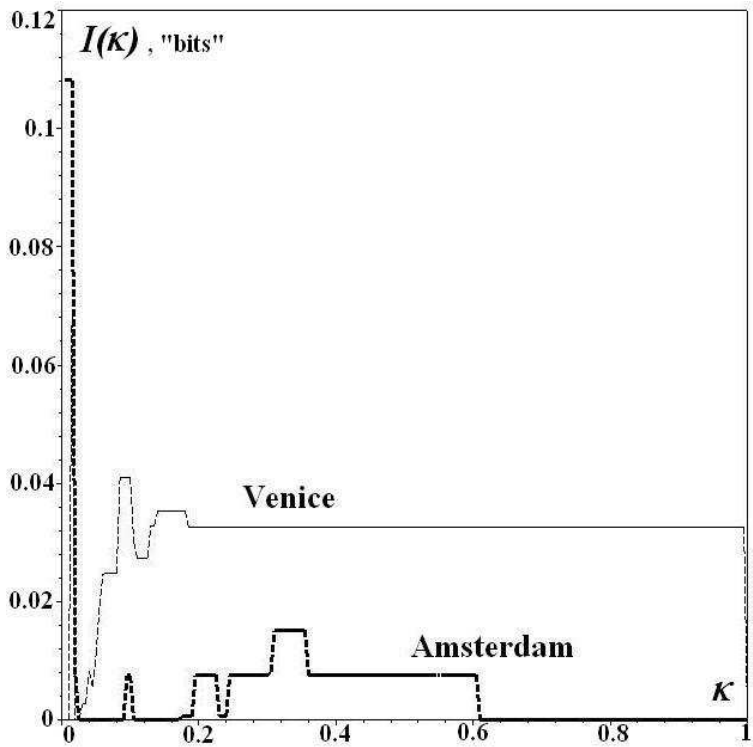, angle= 0,width=10.0cm,height=10cm}
\end{center}
\caption{Quantity of information (bits) encoded by the dynamical
modularities of the Venice and Amsterdam canal networks
 via the threshold of essential correlations, $0\leq\kappa\leq 1$.}
\end{figure}

\newpage

\begin{figure}[ht]
 \noindent
\begin{center}
\epsfig{file=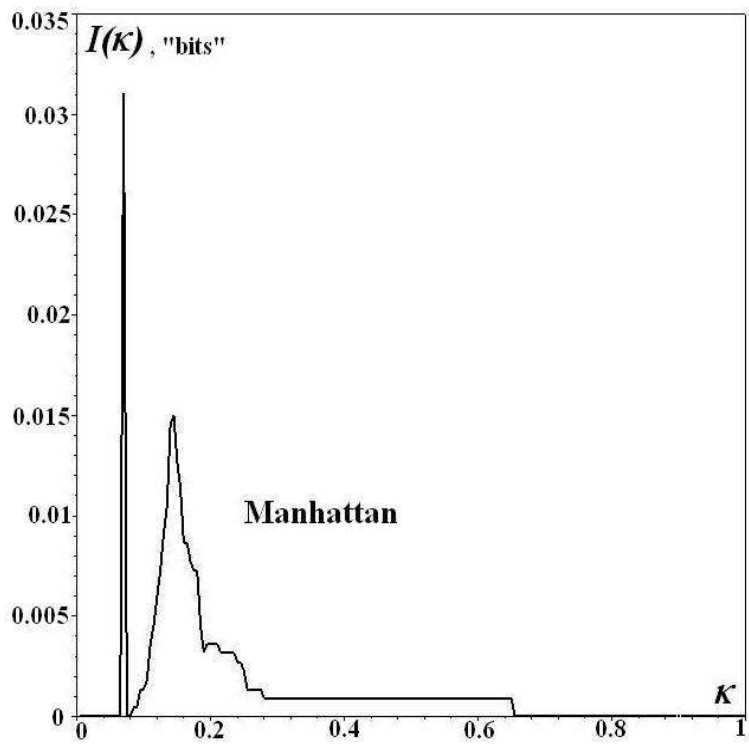, angle= 0,width=10.0cm,height=10cm}
\end{center}
\caption{Quantity of information (bits) encoded by the dynamical
modularities of Manhattan
 via the threshold  of essential correlations  $0\leq\kappa\leq 1$. }
\end{figure}

\newpage

\begin{figure}[ht]
 \noindent
\begin{center}
\epsfig{file=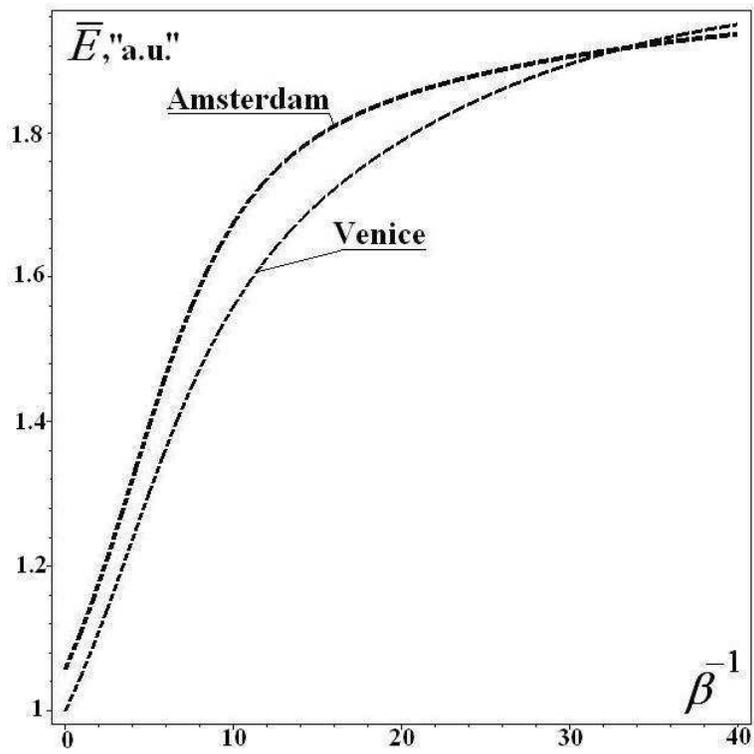, angle= 0,width=10.0cm,height=10cm}
\end{center}
\caption{The grows of "internal energy" in the models of lazy random
 walks with "temperature" $\beta^{-1}$
(the inverse parameter of lazy random walks)
 for the canal networks of Amsterdam and Venice.}
\end{figure}

\newpage

\begin{figure}[ht]
 \noindent
\begin{center}
\epsfig{file=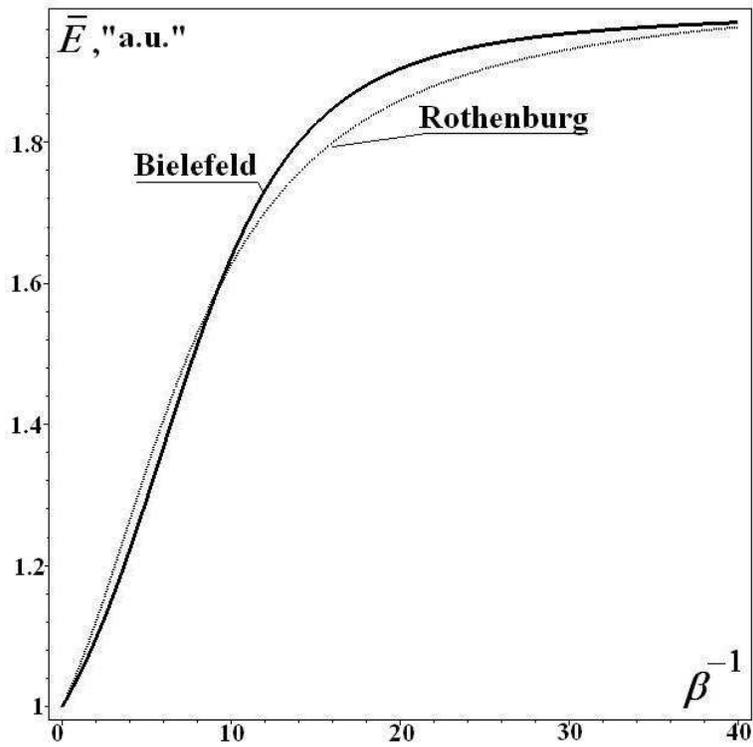, angle= 0,width=10.0cm,height=10cm}
\end{center}
\caption{The grows of "internal energy" in the models of lazy random
walks with "temperature" $\beta^{-1}$
 (the inverse parameter of lazy random walks) for the medieval cities,
  Rothenburg o.d.T. and Bielefeld.}
\end{figure}

\newpage

\begin{figure}[ht]
 \noindent
\begin{center}
\epsfig{file=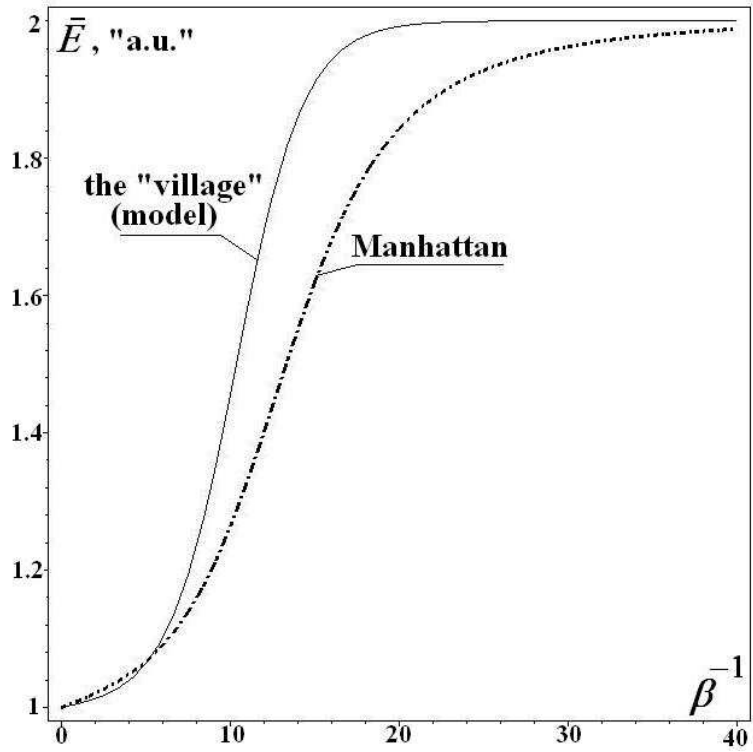, angle= 0,width=10.0cm,height=10cm}
\end{center}
\caption{The grows of "internal energy" in the models of lazy
random walks with "temperature" $\beta^{-1}$ (the inverse
parameter of lazy random walks) for Manhattan and the
 theoretical example of a "village" extended along the only principal street.}
\end{figure}

\newpage

\begin{figure}[ht]
 \noindent
\begin{center}
\epsfig{file=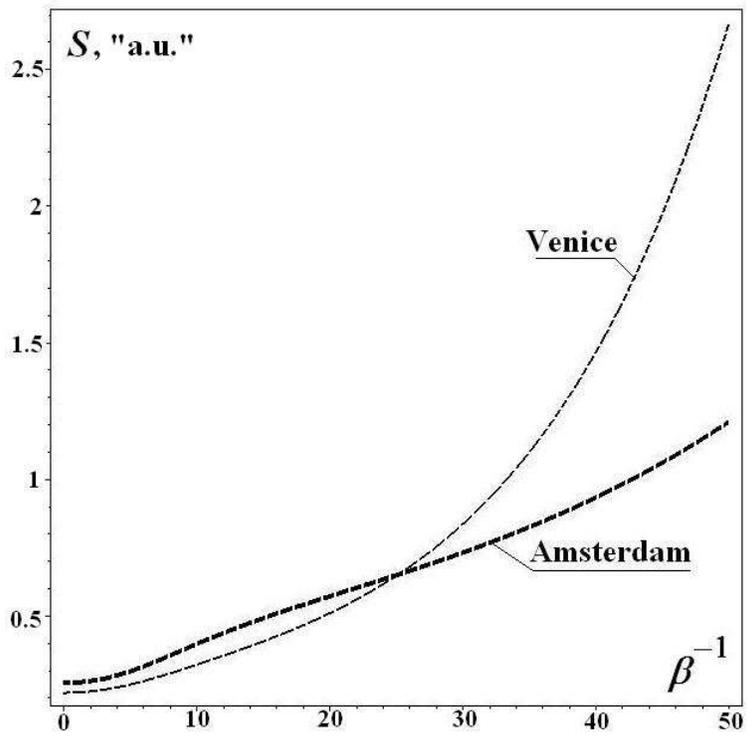, angle= 0,width=10.0cm,height=10cm}
\end{center}
\caption{The entropy curves via the inverse parameter of lazy random
walks, $\beta^{-1},$ for the canal networks of Venice and Amsterdam}
\end{figure}
\newpage

\begin{figure}[ht]
 \noindent
\begin{center}
\epsfig{file=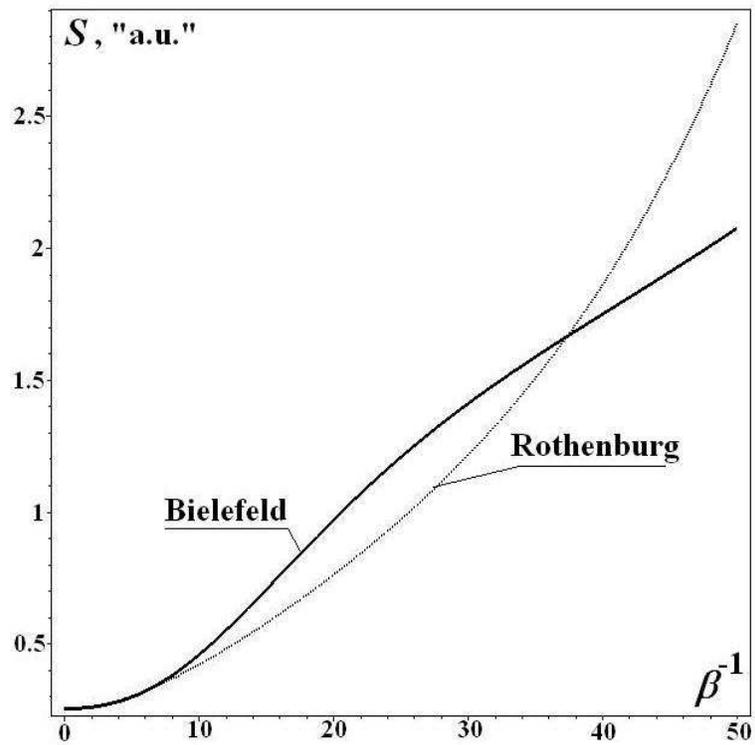, angle= 0,width=10.0cm,height=10cm}
\end{center}
\caption{The entropy curves via the inverse parameter of lazy random
walks, $\beta^{-1}$, for Bielefeld
and Rothenburg o.d.T.}
\end{figure}

\newpage

\begin{figure}[ht]
 \noindent
\begin{center}
\epsfig{file=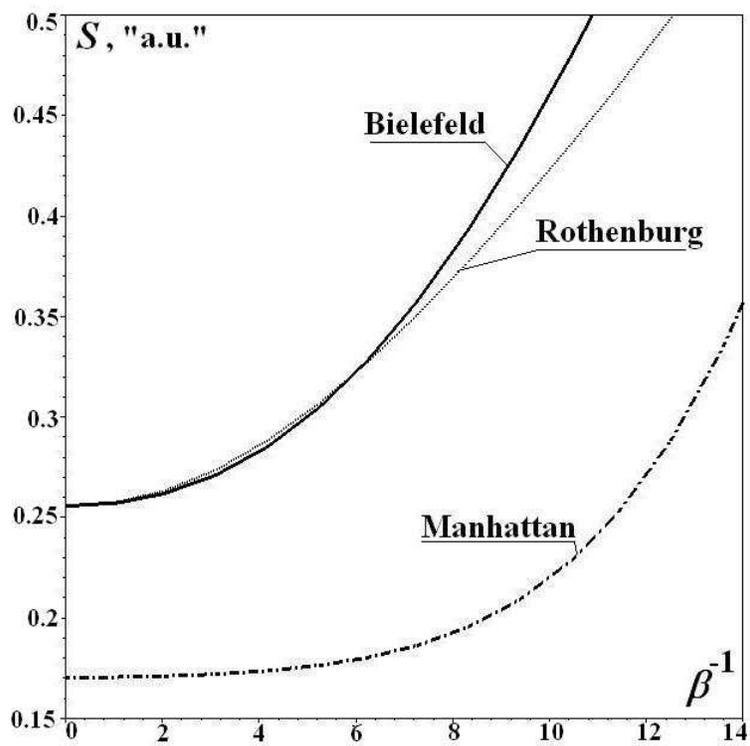, angle= 0,width=10.0cm,height=10cm}
\end{center}
\caption{The entropy curves via the inverse parameter of lazy random
walks, $\beta^{-1}$, for Bielefeld, Rothenburg,
and Manhattan.}
\end{figure}

\newpage

\begin{figure}[ht]
 \noindent
\begin{center}
\epsfig{file=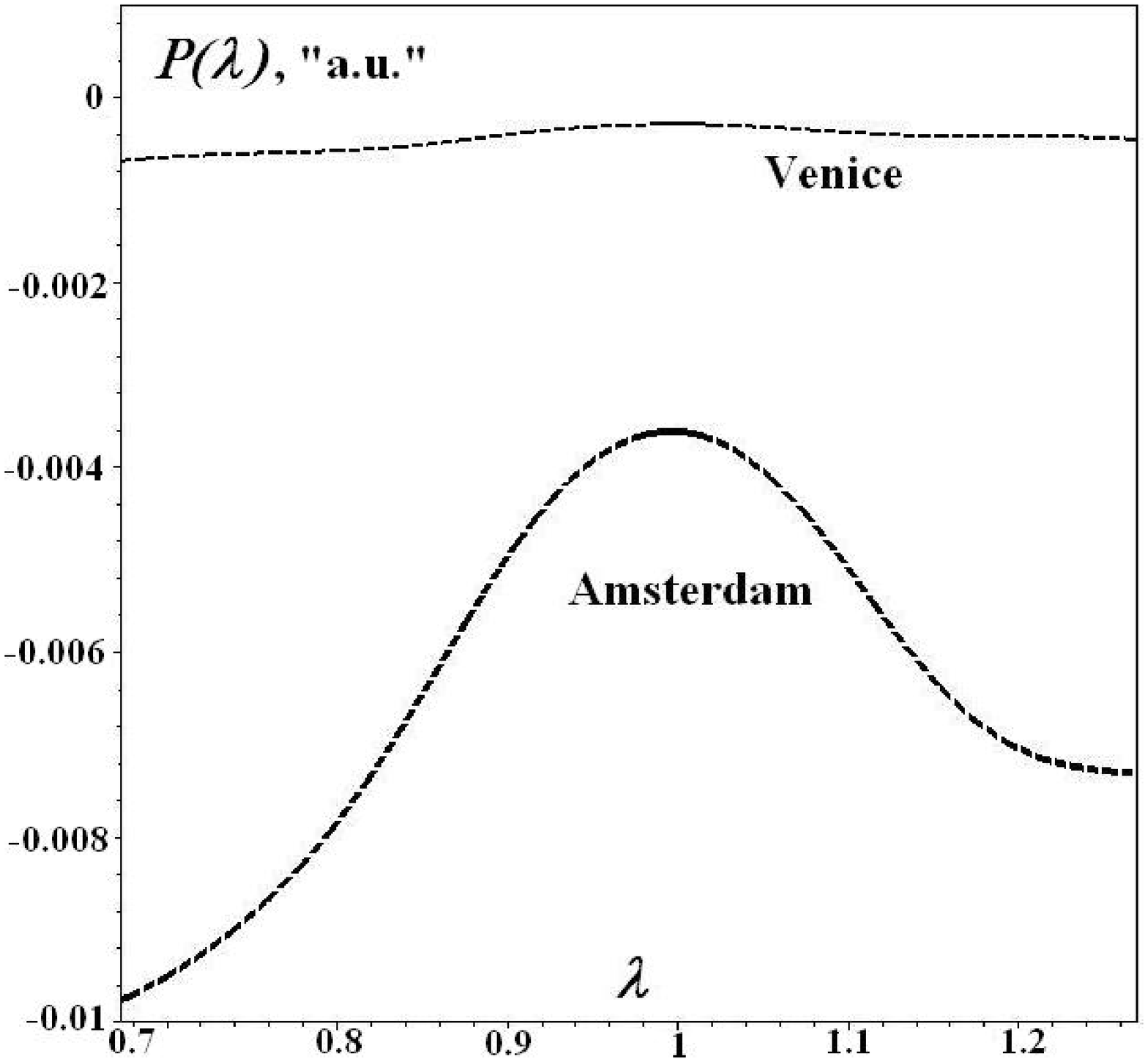, angle= 0,width=10.0cm,height=10cm}
\end{center}
\caption{The comparison of pressure spectra $P(\lambda)$ acting on
the
 flows of  random walkers
with eigenmodes $\lambda$ into Amsterdam and Venice.}
\end{figure}

\newpage

\begin{figure}[ht]
 \noindent
\epsfig{file=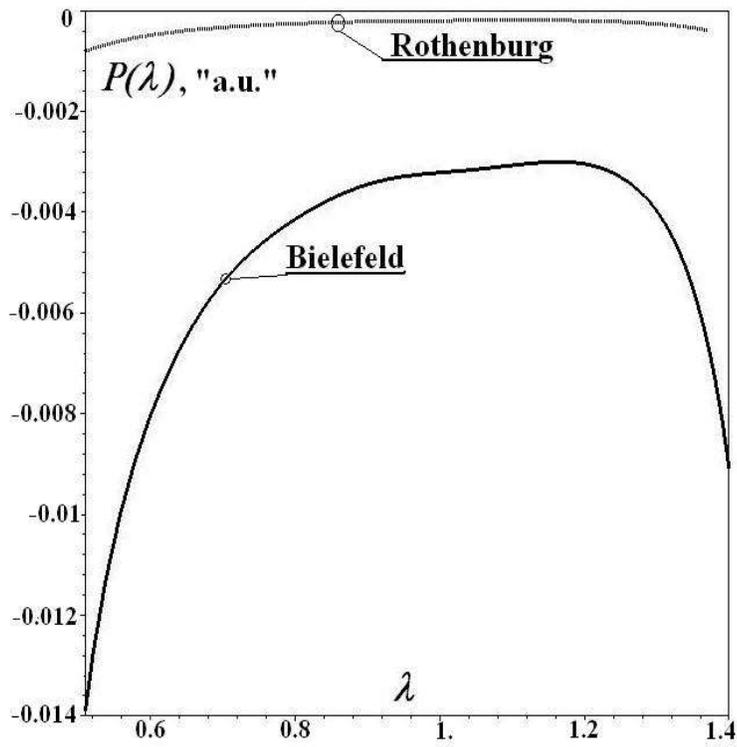, angle= 0,width=10.0cm,height=10cm}
\caption{The comparison of pressure spectra $P(\lambda)$  acting
on the
 flows of  random walkers
with eigenmodes $\lambda$ into Bielefeld and Rothenburg.}
\end{figure}

\clearpage
\newpage

\begin{figure}[ht]
 \noindent
 \epsfig{file=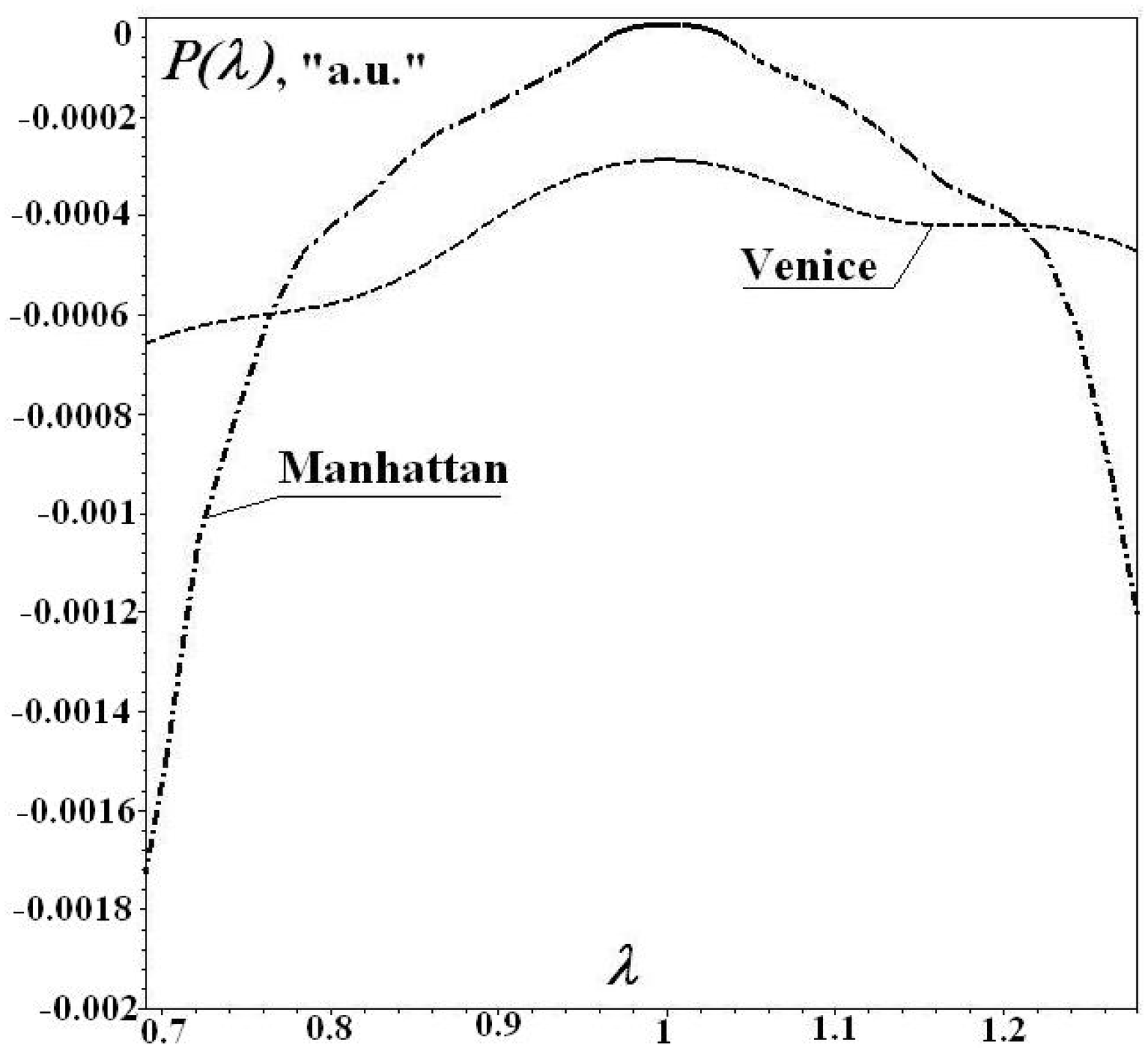, angle= 0,width=10.0cm,height=10cm}
\caption{The comparison of pressure spectra $P(\lambda)$  acting
on the
 flows of  random walkers
with eigenmodes $\lambda$ into Manhattan and Venice.}
\end{figure}


\begin{thebibliography}{000}
\bibitem{Maple}
The 3d-drawing of networks with respect to thier eigenvectors is
used in the standard routines of Maple and Matlab environments.
\end{thebibliography}
\end{document}